\documentclass[lettersize,journal]{IEEEtran}
\usepackage{amsmath,amsfonts}
\usepackage{algorithmic}
\usepackage{algorithm}
\usepackage{array}
\usepackage[caption=false,font=normalsize,labelfont=sf,textfont=sf]{subfig}
\usepackage{textcomp}
\usepackage{stfloats}
\usepackage{url}
\usepackage{verbatim}
\usepackage{graphicx}
\usepackage{amssymb}
\usepackage{cite}
\usepackage{booktabs}
\usepackage{pifont} 
\usepackage{tabularx, makecell, multirow} 
\begin{document}

\title{De-LightSAM: Modality-Decoupled Lightweight SAM for Generalizable Medical Segmentation}

\author{Qing Xu, Jiaxuan Li, Xiangjian He, \IEEEmembership{Senior Member, IEEE}, Chenxin Li, Fiseha B. Tesem, Wenting Duan, Zhen Chen, Rong Qu, \IEEEmembership{Senior Member, IEEE}, Jonathan M. Garibaldi, \IEEEmembership{Fellow, IEEE},  Chang Wen Chen, \IEEEmembership{Fellow, IEEE}  
	\thanks{\quad This work is partially supported by the Yongjiang Technology Innovation Project (2022A-097-G), Zhejiang Department of Transportation General Research and Development Project (2024039), and National Natural Science Foundation of China talent grant (UNNC: B0166). \textit{(Equal contribution: Qing Xu and Jiaxuan Li, Corresponding author: Xiangjian He)}}
        \thanks{\quad Q. Xu, J. Li, X. He, F. B. Tesema and J. M. Garibaldi are with School of Computer Science, University of Nottingham Ningbo China, Ningbo, Zhejiang, China (e-mail: sean.he@nottingham.edu.cn).}
        \thanks{\quad C. Li is with Department of Electronic Engineering, The Chinese University of Hong Kong, Hong Kong SAR (e-mail: chenxinli@link.cuhk.edu.hk).}
        \thanks{\quad W. Duan is with School of School of Engineering \& Physical Sciences, University of Lincoln, Lincoln  LN6 7TS, UK (email: wduan@lincoln.ac.uk).}
        \thanks{\quad Z. Chen is with Hong Kong Institute of Science \& Innovation, Chinese Academy of Sciences, Hong Kong SAR (e-mail: zchen.francis@gmail.com).}
        \thanks{\quad R. Qu is with the School of Computer Science, University of Nottingham, Nottingham NG72RD, UK (email: rong.qu@nottingham.ac.uk).}
        \thanks{\quad C. W. Chen is with The Hong Kong Polytechnic University, Hong Kong (e-mail: changwen.chen@polyu.edu.hk).} 
        }


\markboth{IEEE Transactions on Circuits and Systems for Video Technology}%
{}



\maketitle

\begin{abstract}
The universality of deep neural networks across different modalities and their generalization capabilities to unseen domains play an essential role in medical image segmentation. The recent segment anything model (SAM) has demonstrated strong adaptability across diverse natural scenarios. However, the huge computational costs, demand for manual annotations as prompts and conflict-prone decoding process of SAM degrade its generalization capabilities in medical scenarios. To address these limitations, we propose a modality-decoupled lightweight SAM for domain-generalized medical image segmentation, named De-LightSAM. Specifically, we first devise a lightweight domain-controllable image encoder (DC-Encoder) that produces discriminative visual features for diverse modalities. Further, we introduce the self-patch prompt generator (SP-Generator) to automatically generate high-quality dense prompt embeddings for guiding segmentation decoding. Finally, we design the query-decoupled modality decoder (QM-Decoder) that leverages a one-to-one strategy to provide an independent decoding channel for every modality, preventing mutual knowledge interference of different modalities. Moreover, we design a multi-modal decoupled knowledge distillation (MDKD) strategy to leverage robust common knowledge to complement domain-specific medical feature representations. Extensive experiments indicate that De-LightSAM outperforms state-of-the-arts in diverse medical imaging segmentation tasks, displaying superior modality universality and generalization capabilities. Especially, De-LightSAM uses only 2.0\% parameters compared to SAM-H. The source code is available at \url{https://github.com/xq141839/De-LightSAM}.
\end{abstract}

\begin{IEEEkeywords}
Medical image segmentation, knowledge distillation, domain generalization 
\end{IEEEkeywords}

\section{Introduction}
\label{sec:introduction}

\IEEEPARstart{M}{edical} imaging has made great strides in the last decades, spawning a variety of modalities, such as histopathology imaging, dermoscopy imaging, X-ray imaging, fundus imaging, colonoscopy imaging and ultrasound imaging. They play an important role in determining disease types and grading \cite{li2024focus, zhou2024uncertainty, wang2024frequency}. Traditionally, medical images are analyzed by medical experts, which is time-consuming and occupies substantial healthcare resources. In this challenging background, computer-aided diagnosis is expected to accelerate evaluation time and improve diagnostic efficiency, where pixel-level segmentation of target regions is a key step for quantitative and qualitative assessment, presenting valuable information \cite{wu2023medsa_jinyueming}.

\begin{figure}[!t]
  \centering
  \includegraphics[width=1\linewidth]{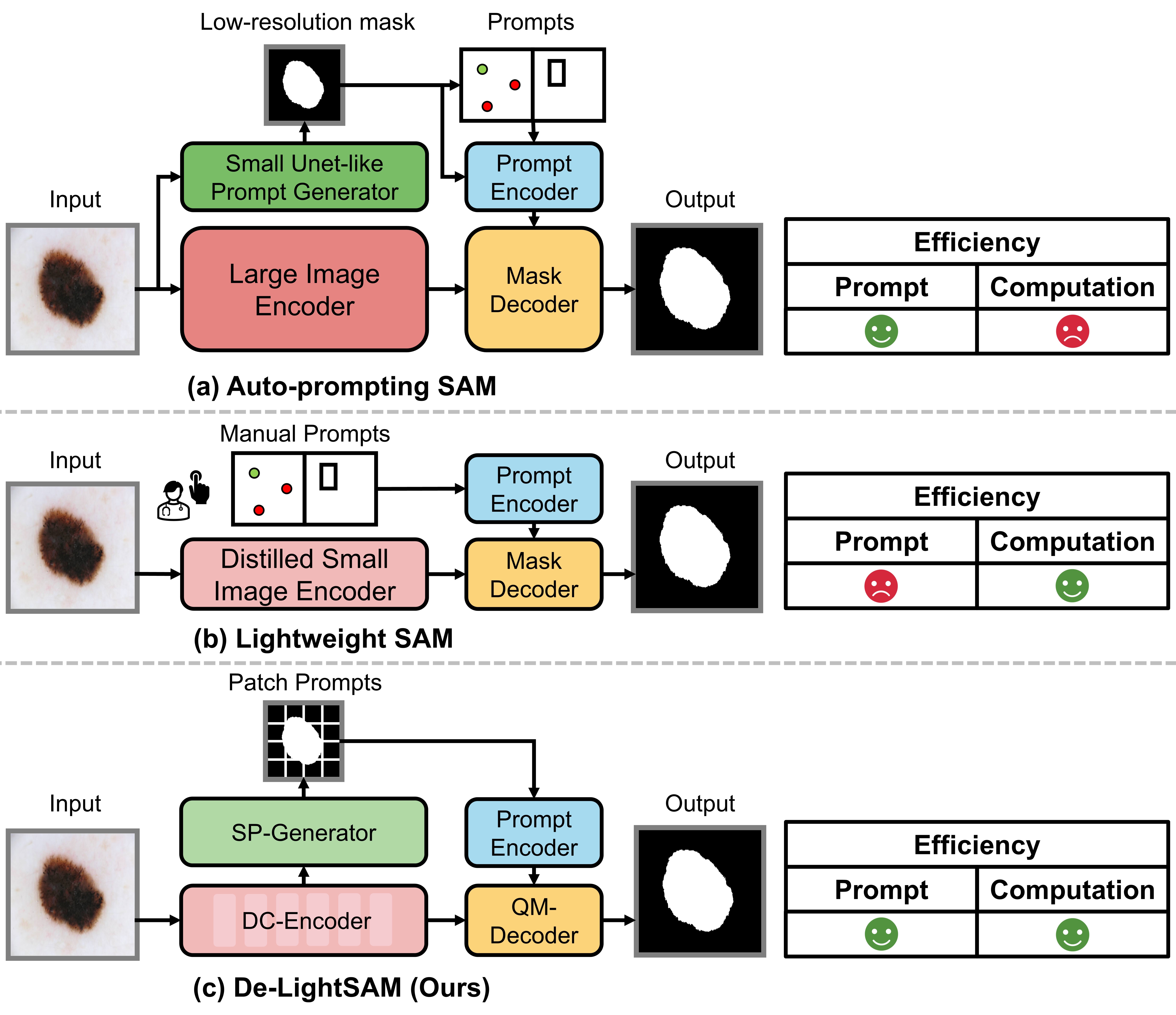}
  \caption{Comparison of our De-LightSAM and existing medical SAM works. (a) Small Unet-like models for automatic prompt generation. (b) Distilled image encoders for lower computation cost. (c) Our De-LightSAM achieves efficient computation cost and auto-prompting generation simultaneously.}
  \label{fig:intro}
\end{figure}

Convolutional neural network (CNN) based U-shape architectures investigate the correlation between the low-level and high-level semantic information in mask prediction \cite{ronneberger2015u, schlemper2019attention,isensee2021nnu,he2023h2former, yun2023spectr, qiao2024effective, li2023erdunet}. Although these methods demonstrate the accurate generation of segmentation masks within known domains, they are difficult to generalize to unseen domains. To address domain generalization (DG) challenges, existing studies utilize multi-task learning, data augmentation and domain synthesis to improve the diversity of model feature representations \cite{nam2024modality, ouyang2022causality, xu2022adversarial}. However, these task-specific models require to be retrained from scratch when facing different modalities due to their limited model capacity. This raises significant challenges in establishing a segmentation model with superior generalization and universality in various medical imaging modalities.

Vision transformer (ViT) \cite{dosovitskiy2020image} leverages the self-attention mechanism to capture long-range dependencies and provides a larger model capacity to overcome the constraints of inductive bias under the supervision of big data. Especially, the recent appearance of segment anything model (SAM) \cite{kirillov2023_sam} has made a significant breakthrough in the field of image segmentation. The superior generalization of SAM in natural images has been sufficiently validated, demonstrating its transferability in a wide range of scenarios through interactive prompts. Existing studies \cite{huang2024segment, ma2024sam_majun_nc} have illustrated the potential of SAM for universal medical image segmentation. Despite these advantages, adapting SAM to clinical scenarios faces three significant obstacles. Firstly, SAM \cite{kirillov2023_sam} mainly depends on manual annotations (e.g., points, boxes) as prompts to guide the segmentation mask generation, leading to labor-intensive operations. Particularly, for the most effortless mode (i.e., single positive point) and simple automatic prompt generation algorithm (i.e., sliding window), recent studies \cite{huang2024segment, ma2024sam_majun_nc} have indicated that they are difficult to perform acceptable results in medical applications due to insufficient or incorrect prompt information. Although current studies \cite{na2024segment, zhang2024uv, shui2025unleashing} include classical small Unet or object detection models, as shown in Fig. \ref{fig:intro}(a), to automatically generate different prompts, they involve significant additional computation cost.

Furthermore, the standard SAM \cite{kirillov2023_sam} contains a large number of parameters. In particular, the image encoder of SAM-H contains 636M parameters. The huge computational costs limit the applicability of SAM in real-world scenarios. Existing methods based on parameter-efficient fine-tuning techniques \cite{wu2023medsa_jinyueming} enable the reduction of learnable parameters during the fine-tuning phase. In addition, to reduce inference cost, many studies \cite{zhang2023faster, xiong2023efficientsam, zhang2024efficientvit} leverage knowledge distillation strategies to transfer the knowledge of SAM to a single small image encoder, as illustrated in Fig. \ref{fig:intro}(b). Meanwhile, they leverage a modality-agnostic decoding query to predict potential segmentation masks for all categories. However, in the field of medical imaging, every vision modality has inherent heterogeneity. This one-to-many strategy is difficult to handle the mutual knowledge interference of different modalities, resulting in the degradation of model generalization capabilities.

To address these limitations, we propose De-LightSAM, a modality-decoupled lightweight SAM framework specifically designed for domain-generalized medical image segmentation, as shown in Fig. \ref{fig:intro}(b) Specifically, we first develop a lightweight domain-controllable image encoder (DC-Encoder) that efficiently captures discriminative visual features across diverse medical modalities while significantly reducing computational overhead. Second, we introduce a self-patch prompt generator (SP-Generator) that eliminates the need for manual annotations by automatically generating high-quality dense prompt embeddings, enabling fully automated segmentation. Third, we design a query-decoupled modality decoder (QM-Decoder) that employs a one-to-one decoding strategy, providing independent channels for each modality to prevent cross-modal knowledge interference and enhance segmentation accuracy. Additionally, we propose a multi-modal decoupled knowledge distillation (MDKD) strategy that effectively transfers both common and domain-specific knowledge from foundation models to our lightweight architecture. Through extensive experiments across multiple medical imaging datasets, we demonstrate that De-LightSAM achieves superior performance compared to state-of-the-art methods while using only 2.0\% of SAM-H's parameters, showcasing exceptional computation efficiency and cross-domain generalization capabilities.

The contributions of this work are summarized as follows. 
\begin{itemize} 
    \item We propose De-LightSAM as an end-to-end framework that addresses real-world medical imaging challenges, including computational constraints, and multi-modal compatibility. De-LightSAM offers a practical pathway in clinical environments with limited computational resources while maintaining high segmentation accuracy across diverse medical imaging scenarios.
    
    \item We devise a DC-Encoder integrated with the MDKD strategy. This approach efficiently extracts discriminative visual features across diverse medical modalities while transferring both common and domain-specific knowledge from foundation models, achieving the significant reduction of computation cost. 
    
    \item We introduce an integrated framework combining an SP-Generator for automatic high-quality prompt generation and a QM-Decoder with a one-to-one decoding strategy. This eliminates manual annotation dependency while providing independent processing channels for each modality to prevent cross-modal interference.
    
    \item We conduct extensive experiments across multiple medical imaging modalities and datasets to systematically evaluate domain generalization capabilities. Our comprehensive analysis demonstrates superior generalization capabilities compared to existing state-of-the-art methods. 
\end{itemize}

\begin{figure*}[t]
  \centering
  \includegraphics[width=1\linewidth]{fig/fig_method.png}
  \caption{The overview of our De-LightSAM framework for domain-generalized medical image segmentation, consisting of DC-Encoder, SP-Generator and QM-Decoder. For ease of understanding, we elaborate on the case of De-LightSAM in skin lesion segmentation. Our De-LightSAM fully exploits a modality-decoupled paradigm to enhance generalization capabilities.}
  \label{fig:method}
  \vspace{-1.0em}
\end{figure*}

\section{Related Work}
\subsection{Generalized Medical Image Segmentation}
The generalization capabilities of deep neural networks have received significant attention in medical image segmentation. The original UNet \cite{ronneberger2015u} reveals great single-domain adaptation but is difficult to generalize unseen domains. Previous studies \cite{schlemper2019attention, zhou2019unet++, nam2024modality, cai2024ultra, zhao2024ultrasound} mainly adopt multi-scale feature fusion to improve the feature representation power of models. In addition, Cheng \textit{et al.} \cite{ouyang2022causality} utilized causality-inspired data augmentation to extend the distribution of the single-source domain during training. Xu \textit{et al.} \cite{xu2022adversarial} proposed an Adversarial Domain Synthesizer (ADS) to synthesize the new domains from the memorized source domain information. As these task-specific methods have limited model capacity, they need to be trained from scratch for each modality. The recent Segment Anything Model (SAM) \cite{kirillov2023_sam} took advantages of its large-scale image encoder and interactive prompts to achieve outstanding zero-shot generalization in natural image segmentation. In medical image segmentation, MedSAM \cite{ma2024sam_majun_nc} and SAMMI \cite{huang2024segment} collected more than 1M public medical images to fully fine-tune SAM with box and point prompts for universal medical image segmentation. However, such methods rapidly increase data and computation costs, which are expensive and impractical in clinical scenarios. To mitigate the reliance on data size and computational resources during transfer learning, parameter-efficient fine-tuning techniques have been introduced in SAM. Specifically, the Adapter has been widely used to integrate into the image encoder of SAM for the refinement of feature representation in medical imaging \cite{wu2023medsa_jinyueming}. SAMed \cite{zhang2023customized_sam_liudong} concatenated Low-Rank Adaptation (LoRA) with self-attention layers of SAM to optimize the feature extraction. While beneficial, these architectures are still based on the huge ViT encoder, which is computationally expensive. In contrast, our approach overcomes this aforementioned challenge and illustrates superior generalization-efficiency trade-offs across a variety of medical imaging modalities. 

Furthermore, SAM \cite{kirillov2023_sam} relying on laborious manual annotations as segmentation prompts seriously reduces its applicability in clinical scenarios. Although it provides a simple sliding window algorithm to automatically generate box and center point prompts from inputs, recent studies \cite{huang2024segment, mazurowski2023segment} have indicated that this approach significantly degrades inference speed and fails to perform satisfactory results in medical image segmentation tasks. To minimize the need for manual annotations, many studies used traditional segmentation networks (e.g., UNet \cite{ronneberger2015u}) to produce low-resolution masks as prompts \cite{zhang2024uv, chen2024sam}. However, such pixel-level small segmentors may generate more error prompts when facing heterogeneous modalities due to their limited model capacity. On the contrary, our De-LightSAM framework automatically generates a set of high-quality patch prompts from its own image embeddings for guiding segmentation decoding, so it eliminates the demand for labour-intensive manual annotations.

\subsection{Knowledge Distillation}
Knowledge distillation (KD) \cite{hinton2015distilling} is a classical method of compressing the size of foundation models, transferring the knowledge from the teacher model to the student model. Generally, hard labels (e.g., category) and soft labels (e.g., probability) are used to supervise the learning of a student model from a teacher model. Yang \textit{et al.} \cite{yang2020knowledge} decoupled the distillation into two stages: representation learning and classification. Subsequently, decoupled knowledge distillation \cite{zhao2022decoupled} was introduced to divide the traditional KD loss into two parts: target class and non-target class knowledge distillation, which enhances the efficiency of knowledge transfer between the teacher and student model. In addition, it has been proven that soft labels are preferable in KD-based medical image segmentation \cite{zhao2023efficient, li2022knowledge}. Recently, transferring the knowledge of SAM to a small model has become a hot research topic. MobileSAM \cite{zhang2023faster} retained the prompt encoder and mask decoder of the standard SAM \cite{kirillov2023_sam}, adopting feature distillation between its image encoder and the TinyViT \cite{wu2022tinyvit}. EdgeSAM \cite{zhou2023edgesam} involved both the prompt encoder and mask decoder in the distillation process to capture the full knowledge embodied in SAM. EfficientSAM \cite{xiong2023efficientsam} utilized masked image pretraining method and reconstruction loss to transfer the knowledge from the image encoder of SAM to a lightweight encoder. However, these feature-coupled distillation methods are challenging to harmonize the feature representation across diverse modalities with inherent heterogeneity, resulting in the degradation of the model generalization capability. On the contrary, our method adopts a feature-level decoupled distillation strategy. The distilled image encoder enables the generation of discriminative features for different medical modalities.

\begin{figure*}[t]
  \centering
  \includegraphics[width=1\linewidth]{fig/fig_method_kd.png}
  \caption{The illustration of our MDKD strategy, including two sub-tasks: common feature distillation and domain-specific feature distillation. We leverage robust common knowledge to complement domain-specific medical feature representations.}
  \label{fig:kd}
  \vspace{-1.0em}
\end{figure*}

\section{Methodology}

\subsection{Overview of De-LightSAM}
In this study, we denote $\mathcal{S}=\{\mathcal{S}^1, \mathcal{S}^2,\cdots, \mathcal{S}^K\}$ as the set of $K$ highly heterogeneous medical vision modalities involved in source domains. Each domain includes image and segmentation mask pairs of $\mathcal{S}^k=\{(x_i,y_i)\}_{i=1}^N$, where $N$ represents the number of pairs within the domain. Our goal is to train a universal segmentation model $\mathcal{F}_{\theta}: x \rightarrow y$ on source domains, which not only performs well across different modalities but also can be directly generalized to an unseen target domain $\mathcal{T}^k$ of each modality.

As illustrated in Fig. \ref{fig:method}, De-LightSAM presents a comprehensive framework for domain-generalized medical image segmentation that operates efficiently across diverse imaging modalities. Given input medical images from the $k$-th modality, our framework processes segmentation through three core components. Firstly, the DC-Encoder extracts discriminative visual representations while maintaining modality-specific characteristics. This lightweight encoder incorporates a modality controller for domain-aware feature learning, optimized through our proposed MDKD strategy from pre-trained foundation models. Then, the SP-Generator automatically transforms the encoded features into high-quality dense prompt embeddings, eliminating the dependency on manual annotations while ensuring comprehensive spatial coverage for accurate segmentation guidance. Finally, the QDMD employs modality-specific decoding channels, where each modality utilizes its dedicated query mechanism to perform independent segmentation inference, thereby preventing cross-modal interference and enhancing modality-specific performance. This end-to-end architecture enables De-LightSAM to achieve superior generalization across unseen medical domains while maintaining computational efficiency.

\subsection{Domain-Controllable Image Encoder}
SAM \cite{kirillov2023_sam} mainly relies on a large-capacity ViT to provide generalized feature representations but is challenged by the huge computation costs, which limit its applications in real-world scenarios. Recent studies \cite{wang2023repvit, zhang2024efficientvit, songa2024sam} mainly aim to transfer the knowledge of the huge image encoder into an independent lightweight architecture. However, different modalities of medical imaging face inherent heterogeneity. Such full-knowledge sharing image encoders are difficult to harmonize the feature representation of the model across diverse modalities, degrading the generalization capabilities. To address this challenge, we propose a DC-Encoder that generates discriminative features for different modalities while maintaining computational efficiency. Given a set of patch embeddings $x^k \in \mathbb{R}^{H \times W \times C}$, where $H$, $W$ and $C_1$ are the height, width and number of channels, our DC-Encoder first employs two MBConvs \cite{howard2019searching} blocks to learn low-level representation efficiently. Specifically, each MBConv block applies a $1 \times 1$ convolution for channel expansion. A $3 \times 3$ depthwise convolution is followed by a $1 \times 1$ projection convolution for effective channel communication. Then, MBConv blocks combine with $L_1$ standard ViT layers to capture long-range dependencies. In particular, we create a modality controller $\mathcal{F}_{\rm{MC}}=\{w^k_{\rm bias}, w^k_{\rm up}, w^k_{\rm down}\}_{k=1}^K$, where $w^k_{\rm bias}$ and $w^k_{\rm MLP}$ represent learnable tokens and MLP layers, respectively. Every $w^k_{\rm bias}$ concatenates with the query $\mathcal{Q}$ branche of the self-attention layer to adapt the attention computation $\mathcal{A}$ of the specific modality:
\begin{equation}
    \mathcal{A} = \mathrm{softmax}(\frac{\mathcal{Q}(x^k \smallfrown w^k_{\rm bias}) \cdot {\mathcal{K}(x^k)}^{T}}{\sqrt{d}})\cdot \mathcal{V}(x^k),
\end{equation}
where $\smallfrown$ is the concatenation operation, $d$ stands for multi-head dimensions and $\cdot$ is the matrix multiplication. After that, the learnable $\{w^k_{\rm up}, w^k_{\rm down}\}$ is combined with the feed-forward network $\mathcal{F}_{\rm FFN}$ of ViT to generate discriminative features $h$ for different modalities:
\begin{equation}
    \textit{h} = x^k + \mathcal{A} + \mathcal{F}_{\rm FFN}(\mathcal{F}_{\rm LN}(\mathcal{A})) + f(w^k_{\rm up}\mathcal{F}_{\rm LN}(\mathcal{A}))w^k_{\rm down},
\end{equation}
where $f(\cdot)$ is the activation layer (e.g., GELU) and $\mathcal{F}_{\rm LN}(\cdot)$ is the LayerNorm operation. In addition, the embedding dimension width $C_1$ and network depth $L_1$ are critical factors affecting computational cost. Previous research \cite{wu2022tinyvit} has demonstrated that wider dimensions facilitate learning of complex representations, while deeper models may lead to overfitting on medical datasets due to limited fully-annotated labels caused by expensive pixel-level annotation costs. To build a lightweight yet effective architecture, our DC-Encoder adopts a balanced design with moderate embedding dimensions and shallow structure (i.e., $C_1=192, L_1=10$). This design choice ensures that Encoder can effectively capture domain-specific features while maintaining computational efficiency.

\subsection{Self-Patch Prompt Generator}
Current medical SAM \cite{huang2024segment, ma2024sam_majun_nc, wu2023medsa_jinyueming, zhang2023customized_sam_liudong} and lightweight SAM \cite{zhang2023faster, wang2023repvit, zhang2024efficientvit, xiong2023efficientsam, zhou2023edgesam, songa2024sam} mainly leverage manual prompts (e.g., points and boxes) to guide the model to provide satisfactory segmentation masks. However, such methods rely on the experience of pathologists in medical scenarios, which are expensive and time-consuming. To eliminate the demand for manual annotations, we propose the SP-Generator that contains a patch generator and prompt encoder to automatically produce a set of high-quality patch prompts to assist segmentation decoding. Specifically, the patch generator $\delta$ consists of $L_2$ convolutional layers for patch merging and $1 \times 1$ convolution $\mathcal{F}_{\rm Conv}^{\rm 1 \times 1}$ for channel compression, where each layer includes $2 \times 2$ convolution $\mathcal{F}_{\rm Conv}^{\rm 2 \times 2}$ with stride 2 followed by LayerNorm. With each additional convolutional layer, the scale of the patch is doubled. After that, we utilize the sigmoid $\sigma$ function to generate probability maps. The self-predicted patch prompts $\hat{p}_{\rm pat}$ is optimized by binary cross-entropy loss since this prediction can be considered as a global classification task:
\begin{equation} 
L_{\rm pat}= -p_{\rm pat}\log (\hat{p}_{\rm pat})
 +(1-p_{\rm pat}) \log (1-\hat{p}_{\rm pat}),
\end{equation}
where $p_{\rm pat}$ is the target patch generated by the maxpooled ground truth. We then respectively use the interpolation method for memory-efficient upsampling $\hat{p}_{\rm pat}$ and $1 \times 1$ convolution to align the dimension with the image embedding $h$. They constitute the prompt encoder. During inference, the computation of dense prompt embeddings $p_{\rm dense}$ can be formulated as:
\begin{equation}
    p_{\rm dense} = \mathcal{F}_{\rm Conv}^{\rm 1 \times 1}(\mathcal{F}_{\rm inter}(\sigma(\mathcal{F}_{\rm Conv}^{\rm 1 \times 1}(\mathcal{F}_{\rm LN}(\mathcal{F}_{\rm Conv}^{\rm 2 \times 2}(h))))))).
\end{equation}
As the image embedding $\textit{h}$ contains rich semantic information and the target area of many medical images is much smaller than the background, we place the patch generator on $\textit{h}$. The patch prompt is essentially an inductive prediction, reducing the complexity of predicting the final segmentation mask by providing the summarized semantic information to the decoder. In this way, the SP-Generator module automatically produces a set of high-quality dense prompts to guide the prediction of segmentation masks, improving the applicability of De-LightSAM in clinical scenarios.

\subsection{Query-Decoupled Modality Decoder}
The mask decoder of SAM \cite{kirillov2023_sam} utilizes modality-agnostic query tokens to handle all segmentation tasks in natural images. However, this is not optimal for medical image segmentation. As there exists inherent heterogeneity in various medical imaging modalities, such common prediction channels suffer from decoding conflicts, degrading the generalization capabilities of the model. To address this problem, we propose the QM-Decoder for our De-LightSAM framework.
Concretely, we set $K$ query tokens $\{q^k\}^K_{k=1}$, where $q^k \in \mathbb{R}^{1\times C_2}$, to customize the private workflow for each Modality. We adopt a class-fixed assignment algorithm where each query token corresponds to one modality category. Given the image embeddings $\textit{h}$ and prompt embeddings $\textit{p}$, we first update the mask query using the self-attention operation and then conduct cross-attention with the fusion of $\textit{h}$ and $p_{\rm dense}$: $h \gets h \oplus p_{\rm dense}$, where $\oplus$ is an element-wise addition operation:
\begin{equation}
\begin{aligned}
    q^k \gets \mathrm{softmax}(\frac{\mathcal{Q}(q^k) \cdot  \mathcal{K}(q^k)^{T}}{\sqrt{d}})\cdot \mathcal{V}(q^k), \\
\end{aligned}
\end{equation}

\begin{equation}
\begin{aligned}
    h' = \mathrm{softmax}(\frac{\mathcal{Q}(h \oplus \Omega) \cdot  \mathcal{K}(q^k)^{T}}{\sqrt{d}})\cdot \mathcal{V}(h) \oplus h, \\
\end{aligned}
\end{equation}
where $h'$ is the decoding embedding, $\Omega$ stands for the corresponding positional encodings. Similar to SAM \cite{kirillov2023_sam}, such operations are iterated twice. To predict the segmentation mask $\hat{y}^k$, we perform: 
\begin{equation} 
\hat{y}^k = \mathcal{F}_{\rm inter}(\sigma(\mathcal{F}^{2 \times 2}_{\rm trans}(\mathcal{F}^{2 \times 2}_{\rm trans}(h')) \cdot \mathcal{F}^k_{\rm MLP}(h')),
\end{equation}
where $\mathcal{F}^{2 \times 2}_{\rm trans}$ represents two $2 \times 2$ transpose convolutions, upsampling $h'$ by 4$\times$. $k$-th MLP $\mathcal{F}^k_{\rm MLP}$ aligns the channel with upscaled $\mathcal{H}$ and contains the decoding information of corresponding modalities. The sigmoid operation, followed by an interpolation function, is used to recover the size of the original masks. The quality of predicted masks is supervised by a combination of focal loss and dice loss: 
\begin{equation} 
    L_{\rm seg} = L_{\rm focal} + L_{\rm dice}.
\end{equation} 
Consequently, the proposed QM-Decoder provides an independent decoding process for each modality, which avoids the conflicting inherent heterogeneity of different modalities, improving the generalization capabilities of our De-LightSAM.

\subsection{Multi-Modal Decoupled Knowledge Distillation}
Knowledge distillation has emerged as a prominent technique for transferring knowledge from large foundation models to lightweight architectures while maintaining competitive performance. To compress the SAM size, recent studies \cite{zhou2023edgesam, xiong2023efficientsam, zhang2023faster} usually adopt the same modality as the pretraining during the distillation stage. However, the scarcity of medical imaging data poses additional challenges for effective knowledge distillation. Medical datasets are inherently limited in scale due to privacy concerns, expensive annotation costs, and the specialized expertise required for accurate labeling. This data scarcity leads to insufficient sample diversity for robust feature learning, causing the student model to overfit to the limited training samples and fail to generalize to unseen medical domains. To address this issue, we propose an MDKD strategy to provide robust foundational knowledge that can complement domain-specific medical features. Specifically, we construct a multi-modal teacher model that involves two complementary components: (1) a natural image encoder $\mathcal{F}_{\rm{ViT\text{-}SAM}}$ of SAM \cite{kirillov2023_sam} to provide common knowledge as this image encoder is pretrained on a large-scale dataset containing a wide variety of diversity and (2) a medical SAM image encoder $\mathcal{F}_{\rm{ViT\text{-}Med}}$ to provide domain-specific knowledge. To obtain $\mathcal{F}_{\rm{ViT\text{-}Med}}$, we integrate an adapter-based SAM encoder with our SP-Generator and QM-Decoder, adopting the parameter-efficient fine-tuning technique \cite{houlsby2019parameter} to achieve adaptation on source domains. In particular, to reduce the number of learnable parameters, we only insert adapters in FFN layers as they determine the feature generation. Drawing inspiration from the divide-and-conquer algorithm, our proposed MDKD method decouples the KD process into two sub-tasks: common feature distillation and domain-specific feature distillation. During common feature distillation, we aim to transfer general visual representation knowledge from $\mathcal{F}_{\rm{ViT\text{-}SAM}}$ to the backbone components of DC-Encoder (denoted as $\mathcal{F}_{\rm{DCE}}$) without involving the modality controller. This process ensures that our lightweight encoder can capture fundamental visual patterns and structural information that are universally applicable across different imaging modalities. Moreover, we distill the specialized knowledge of each medical modality from $\mathcal{F}_{\rm{ViT\text{-}Med}}$ to the modality controller $\mathcal{F}_{\rm{DCE}}^{\rm MC}$ of our DC-Encoder. This targeted distillation enables the modality controller to learn domain-specific characteristics and adapt the shared backbone features to the unique requirements of each medical imaging modality. By separating this process from common feature distillation, we ensure that modality-specific adaptations do not interfere with the general visual representation capabilities. The loss function of MDKD is defined by:
\begin{equation}
\begin{aligned}
    L_{\rm MDKD} &= \big|\big|\mathcal{F}_{\rm{ViT\text{-}SAM}}(\textbf{\textit{x}}_{\rm nat}) - \mathcal{F}_{\rm{DCE}}(\textbf{\textit{x}}_{\rm nat})\big|\big|^2_2 \\
                &+ \sum^{K}_{k=1} \big|\big|\mathcal{F}_{\rm{ViT\text{-}Med}}(\textbf{\textit{x}}^k) - \mathcal{F}^{\rm MC}_{\rm{DCE}}(\textbf{\textit{x}}^k)\big|\big|^2_2,
\end{aligned}
\end{equation}
where $\textbf{\textit{x}}_{\rm nat}$ stands for the $1\%$ natural images of the SA-1B dataset \cite{kirillov2023_sam} and $\textbf{\textit{x}}^k$ is the medical image collected from source domains $\mathcal{S}^k$. Overall, our MDKD method incorporates large-scale natural image data to provide robust common knowledge, leading to superior segmentation performance and enhanced domain generalization capabilities.

 \begin{table}[!t]
    \centering
    \small
    \caption{Details of the source domains used in our experiments.}\label{tab:d1}
    {\scalebox{0.93}{
    \begin{tabular}{ccccc}
    \toprule
    No. &  Dataset &  Modality & Resolution & Images \\
    \midrule
    $\mathcal{S}^1$ & ISIC-2018 \cite{tschandl2018ham10000, codella2019skin} & Dermoscopy & Variable & 3694\\
    $\mathcal{S}^2$ & PCXA \cite{candemir2013lung} & X-ray & Variable & 704\\
    $\mathcal{S}^3$ & DRIVE \cite{staal2004ridge} & Fundus & 584 $\times$ 565 & 40\\
    $\mathcal{S}^4$ & CVC-ClinicDB \cite{bernal2015wm} & Colonoscopy & 384 $\times$ 288 & 612\\
    $\mathcal{S}^5$ & UDIAT \cite{yap2017automated} & Ultrasound & Variable & 163\\
    $\mathcal{S}^6$ & DSB-2018 \cite{caicedo2019nucleus} & Microscopy & Variable & 670\\
    \bottomrule
    \end{tabular}}}
\end{table}

\begin{table}[!t]
    \centering
    \small
    \caption{Details of the unseen domains used in our experiments.}\label{tab:d2}
    {\scalebox{0.93}{
    \begin{tabular}{ccccc}
    \toprule
    No. &  Dataset &  Modality & Resolution & Images \\
    \midrule
    $\mathcal{T}^1$ & PH2 \cite{mendoncca2013ph} & Dermoscopy & 767 $\times$ 576 & 200\\
    $\mathcal{T}^2$ & NIHCXR \cite{tang2019xlsor} & X-ray & 512 $\times$ 512 & 100\\
    $\mathcal{T}^3$ & STARE \cite{hoover2003locating} & Fundus & 700 $\times$ 605 & 20\\
    $\mathcal{T}^4$ & CVC-ColonDB \cite{vazquez2017benchmark} & Colonoscopy & 574 $\times$ 500 & 380\\
    $\mathcal{T}^5$ & BUSI \cite{al2020dataset} & Ultrasound & Variable & 780\\
    $\mathcal{T}^6$ & TNBC \cite{naylor2018segmentation} & Microscopy & 512 $\times$ 512 & 50\\
    \bottomrule
    \end{tabular}}}
    \vspace{-1.0em}
\end{table}

\begin{table*}[!t]
    \centering
    \small
    \caption{Comparison with State-of-the-art Frameworks in Universal Medical Image Segmentation (source domains).}
    {\scalebox{0.87}{
    \begin{tabular}{lcccccccccccccccccc}
    \toprule 
    \multirow{2}{*}{Methods} & Manual & \multicolumn{2}{c}{$\mathcal{S}^1$} &  \multicolumn{2}{c}{$\mathcal{S}^2$}  & \multicolumn{2}{c}{$\mathcal{S}^3$} &  \multicolumn{2}{c}{$\mathcal{S}^4$}  & \multicolumn{2}{c}{$\mathcal{S}^5$}  & \multicolumn{2}{c}{$\mathcal{S}^6$}\\
    \cline{3-14}
     & Prompt & Dice $\uparrow$ & HD $\downarrow$  & Dice $\uparrow$ & HD $\downarrow$ & Dice $\uparrow$ & HD $\downarrow$ & Dice $\uparrow$ & HD $\downarrow$ & Dice $\uparrow$ & HD $\downarrow$ & Dice $\uparrow$ & HD $\downarrow$ \\
    \midrule
    U-Net \cite{ronneberger2015u} & \multirow{9}{*}{\ding{56}} & 82.87 & 180.90 & 93.85 & 103.51 & 79.13 & 68.22 & 83.91 & 130.96 & 69.24 & 131.62 & 88.16 & 130.28 \\
    U-Net++ \cite{zhou2019unet++} & & 82.69 & 175.04 & 95.31 & 75.84 & 80.61 & 65.03 & 85.77 & 152.96 & 72.50 & 137.84 & 90.48 & 112.12 \\
    Att-UNet \cite{schlemper2019attention} & & 83.97 & 170.78 & 95.80 & 66.16 & 80.72 & 66.48 & 86.90 & 156.72 & 71.02 & 107.70 & 91.12 & 113.64 \\
    nnUNet \cite{isensee2021nnu} & & 84.96 & 126.19 & 96.06 & 68.24 & 81.71 & 64.16 & 88.38 & 127.28 & 75.22 & 119.66 & 91.61 & 121.16 \\
    H2Former \cite{he2023h2former} & & 82.12 & 191.39 & 95.54 & 75.44 & 81.46 & 65.09 & 84.66 & 142.08 & 70.30 & 110.98 & 90.17 & 117.69\\
    TransUNet \cite{chen2024transunet} & & 84.28 & 134.80 & 96.27 & 56.85 & 81.68 & 64.91 & 86.00 & 151.14 & 71.07 & 123.80 & 90.03 & 109.56\\
    ADS \cite{xu2022adversarial} & & 84.14 & 172.84 & 94.89 & 86.68 & 80.48 & 68.49 & 87.70 & 117.74 & 72.55 & 136.08 & 90.32 & 115.24\\
    CIAug \cite{ouyang2022causality} & & 83.91 & 141.07 & 95.82 & 73.18 & 80.45 & 65.72 & 87.69 & 106.68 & 71.78 & 134.84 & 90.58 & 113.72 \\
    MADGNet \cite{nam2024modality} & & 85.02 & 131.84 & 96.22 & 67.41 & 81.89 & 64.73 & 88.20 & 107.16 & 72.75 & 131.24 & 91.38 & 98.04 \\
    \midrule
    MobileSAM \cite{zhang2023faster} & \multirow{6}{*}{Point} & 87.97 & 105.34 & 96.25 & 50.94 & 69.31 & 94.41 & 81.83 & 86.46 & 66.48 & 107.78 & 87.42 & 131.00\\
    RepViT-SAM \cite{wang2023repvit} & & 88.00 & 106.75 & 96.07 & 52.71 & 67.76 & 98.66 & 81.81 & 154.65 & 68.38 & 103.09 & 88.81 & 127.17\\
    EfficientViT-SAM \cite{zhang2024efficientvit} & & 88.49 & 103.61 & 96.43 & 49.07 & 78.16 & 77.16 & 85.16 & 102.72 & 74.71 & 113.18 & 89.37 & 116.42 \\
    EfficientSAM \cite{xiong2023efficientsam} & & 87.11 & 108.12 & 96.40 & 49.77 & 76.32 & 79.41 & 82.81 & 96.35 & 71.17 & 113.57 & 88.41 & 129.08\\
    EdgeSAM \cite{zhou2023edgesam} & & 88.10 & 100.92 & 96.18 & 51.60 & 68.04 & 92.40 & 81.76 & 105.06 & 67.64 & 105.51 & 87.35 & 113.46\\
    SAM-Lightening \cite{songa2024sam} &  & 88.28 & 101.64 & 96.41 & 50.35 & 74.84 & 87.32 & 83.70 & 97.79 & 73.07 & 129.53 & 89.18 & 111.67\\
    \midrule
    MobileSAM \cite{zhang2023faster} & \multirow{6}{*}{\ding{56}} & 86.19 & 168.48 & 94.75 & 175.50 & 25.56 & 254.24 & 77.29 & 284.32 & 61.51 & 342.22 & 61.49 & 307.44\\
    RepViT-SAM \cite{wang2023repvit} & & 85.73 & 157.66 & 94.27 & 170.27 & 25.10 & 263.76 & 78.75 & 296.36 & 61.62 & 440.75 & 60.54 & 287.63\\
    EfficientViT-SAM \cite{zhang2024efficientvit} & & 87.18 & 151.29 & 96.13 & 128.60 & 25.95 & 252.39 & 82.48 & 317.51 & 65.44 & 354.98 & 66.32 & 292.69 \\
    EfficientSAM \cite{xiong2023efficientsam} & & 87.02 & 162.45 & 94.89 & 169.20 & 25.84 & 261.51 & 78.54 & 305.36 & 62.05 & 430.18 & 60.56 & 336.55\\
    EdgeSAM \cite{zhou2023edgesam} & & 85.86 & 153.97 & 94.52 & 183.37 & 25.62 & 257.85 & 76.41 & 328.79 & 60.24 & 400.69 & 59.68 & 357.20\\
    SAM-Lightening \cite{songa2024sam} & & 86.99 & 165.80 & 95.83 & 115.19 & 25.52 & 261.25 & 80.14 & 292.33 & 62.59 & 376.55 & 62.68 & 306.66\\
    \midrule 
    De-LightSAM & \ding{56} & \textbf{88.52} & \textbf{92.42} & \textbf{96.83} & \textbf{40.68} & \textbf{82.42} & \textbf{62.64} & \textbf{92.93} & \textbf{56.32} & \textbf{85.28} & \textbf{82.32} & \textbf{92.24} & \textbf{85.93}\\
    \bottomrule
    \end{tabular}}}
    \label{tab:1}
    \vspace{-1.0em}
\end{table*}

\begin{figure*}[!t]
  \centering
  \includegraphics[width=0.95\linewidth]{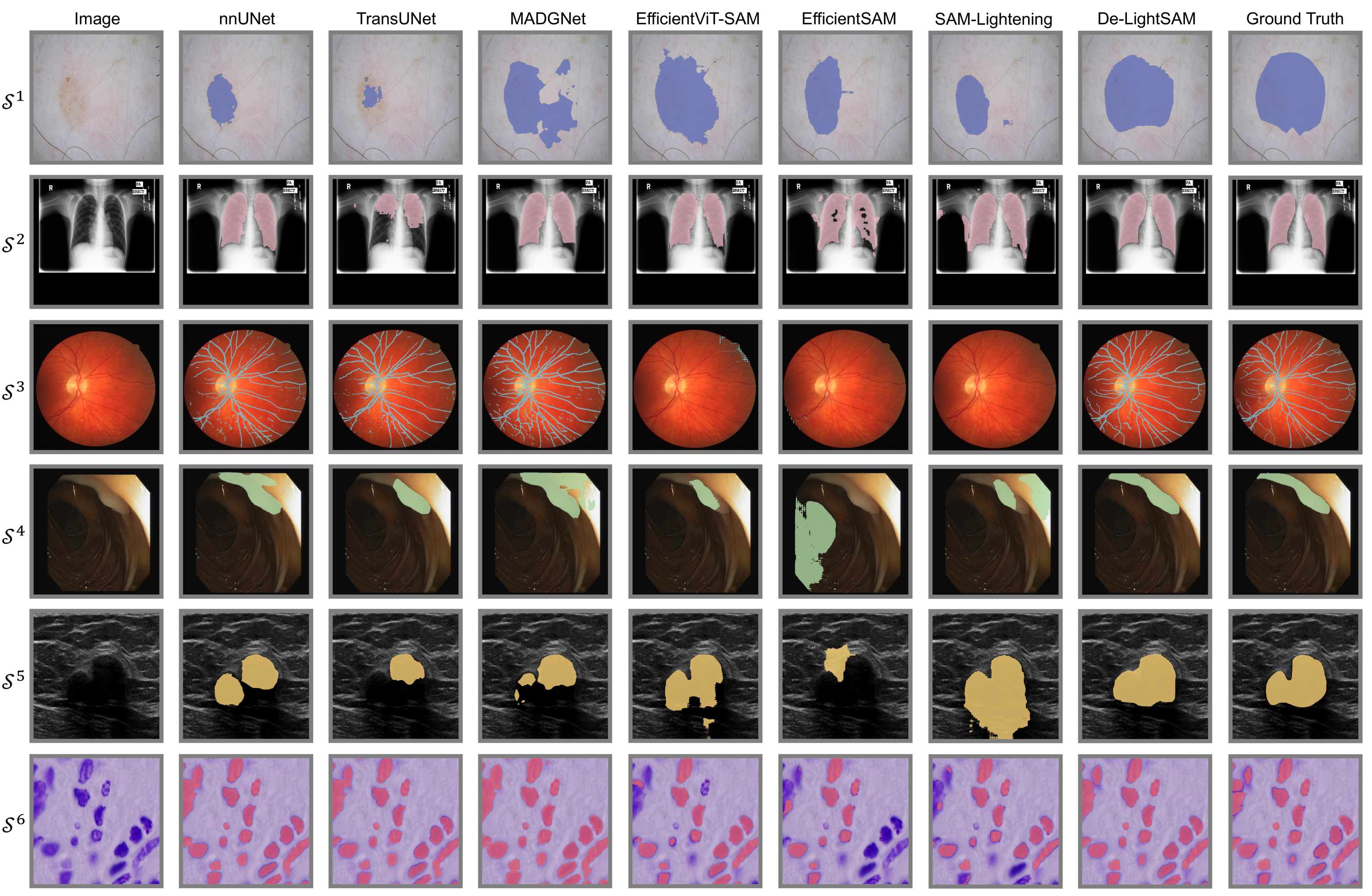}
  \vspace{-1.0em}
  \caption{Qualitative comparison with SOTA task-specific models and lightweight SAM frameworks on medical image segmentation (\textit{source domain}).}
  \label{fig:visual}
  \vspace{-1.0em}
\end{figure*}

\begin{table*}[!t]
    \centering
    \small
    \caption{Comparison with State-of-the-art Frameworks in Domain-Generalized Medical Image Segmentation (unseen domains).}
    {\scalebox{0.87}{
    \begin{tabular}{lcccccccccccccccccc}
    \toprule 
    \multirow{2}{*}{Methods} & Manual & \multicolumn{2}{c}{$\mathcal{S}^1 \rightarrow \mathcal{T}^1$} &  \multicolumn{2}{c}{$\mathcal{S}^2 \rightarrow \mathcal{T}^2$}  & \multicolumn{2}{c}{$\mathcal{S}^3 \rightarrow \mathcal{T}^3$} &  \multicolumn{2}{c}{$\mathcal{S}^4 \rightarrow \mathcal{T}^4$}  & \multicolumn{2}{c}{$\mathcal{S}^5 \rightarrow \mathcal{T}^5$}  & \multicolumn{2}{c}{$\mathcal{S}^6 \rightarrow \mathcal{T}^6$}\\
    \cline{3-14}
     & Prompt & Dice $\uparrow$ & HD $\downarrow$  & Dice $\uparrow$ & HD $\downarrow$ & Dice $\uparrow$ & HD $\downarrow$ & Dice $\uparrow$ & HD $\downarrow$ & Dice $\uparrow$ & HD $\downarrow$ & Dice $\uparrow$ & HD $\downarrow$ \\
    \midrule
    U-Net \cite{ronneberger2015u} & \multirow{9}{*}{\ding{56}} & 87.00 & 140.38 & 69.98 & 223.70 & 61.21 & 174.97 & 32.60 & 461.64 & 39.17 & 336.98 & 46.57 & 295.62\\
    U-Net++ \cite{zhou2019unet++} & & 87.87 & 110.08 & 71.29 & 222.65 & 62.79 & 180.01 & 36.43 & 466.92 & 41.30 & 343.66 & 47.77 & 281.42\\
    Att-UNet \cite{schlemper2019attention} & & 88.66 & 126.35 & 72.41 & 234.26 & 65.18 & 118.10 & 35.56 & 420.60 & 41.89 & 357.08 & 48.70 & 280.56\\
    nnUNet \cite{isensee2021nnu} & & 89.56 & 105.53 & 74.48 & 211.68 & 65.00 & 107.82 & 36.00 & 490.92 & 43.87 & 269.04 & 49.19 & 283.67\\
    H2Former \cite{he2023h2former} & & 87.79 & 146.24 & 73.34 & 219.59 & 65.68 & 115.71 & 34.72 & 480.78 & 42.46 & 275.80 & 53.86 & 294.98\\
    TransUNet \cite{chen2024transunet} & & 89.26 & 108.48 & 80.11 & 193.36 & 66.23 & 112.06 & 42.79 & 339.14 & 44.84 & 267.26 & 54.22 & 282.64\\
    ADS \cite{xu2022adversarial} & & 87.83 & 129.46 & 76.92 & 180.72 & 62.45 & 172.16 & 37.36 & 454.70 & 43.19 & 276.30 & 51.06 & 281.34\\
    CIAug \cite{ouyang2022causality} & & 88.44 & 127.62 & 68.22 & 257.36 & 65.65 & 120.87 & 39.07 & 387.02 & 41.50 & 267.40 & 53.92 & 286.23\\
    MADGNet \cite{nam2024modality} & & 89.71 & 96.86 & 84.11 & 168.58 & 66.88 & 119.41 & 44.32 & 365.84 & 44.91 & 264.61 & 59.29 & 278.56\\
    \midrule
    MobileSAM \cite{zhang2023faster} & \multirow{6}{*}{Point} & 90.37 & 87.28 & 89.14 & 140.02 & 54.10 & 160.16 & 32.77 & 399.29 & 38.68 & 310.52 & 16.17 & 452.18\\
    RepViT-SAM \cite{wang2023repvit} & & 90.63 & 84.74 & 88.17 & 148.25 & 55.72 & 133.82 & 27.76 & 381.39 & 33.58 & 301.85 & 15.20 & 438.64\\
    EfficientViT-SAM \cite{zhang2024efficientvit} & & 91.14 & 85.05 & 89.73 & 145.84 & 72.12 & 116.89 & 61.67 & 179.99 & 58.58 & 183.63 & 34.24 & 332.94 \\
    EfficientSAM \cite{xiong2023efficientsam} & & 90.80 & 89.34 & 89.00 & 150.07 & 69.20 & 98.05 & 56.50 & 218.22 & 52.95 & 233.98 & 25.14 & 358.13\\
    EdgeSAM \cite{zhou2023edgesam} & & 90.38 & 86.32 & 88.01 & 147.62 & 56.03 & 136.48 & 28.17 & 433.08 & 37.51 & 291.65 & 12.11 & 501.14\\
    SAM-Lightening \cite{songa2024sam} &  & 90.85 & 89.96 & 89.63 & 144.84 & 67.38 & 98.85 & 58.12 & 210.20 & 54.75 & 244.37 & 23.69 & 423.15\\
    \midrule
    MobileSAM \cite{zhang2023faster} & \multirow{6}{*}{\ding{56}} & 85.61 & 304.75 & 87.60 & 277.89 & 2.23 & 253.99 & 20.90 & 470.94 & 31.03 & 342.10 & 6.44 & 375.52\\
    RepViT-SAM \cite{wang2023repvit} & & 84.69 & 283.65 & 86.18 & 290.07 & 2.30 & 283.22 & 17.64 & 491.23 & 26.64 & 409.92 & 5.96 & 383.02\\
    EfficientViT-SAM \cite{zhang2024efficientvit} & & 89.29 & 177.12 & 89.33 & 287.08 & 9.72 & 256.58 & 58.77 & 357.63 & 46.08 & 501.00 & 14.85 & 442.26 \\
    EfficientSAM \cite{xiong2023efficientsam} & & 88.44 & 281.54 & 88.24 & 281.52 & 6.75 & 298.82 & 46.99 & 474.12 & 39.44 & 468.81 & 12.81 & 419.71\\
    EdgeSAM \cite{zhou2023edgesam} & & 84.31 & 334.93 & 87.07 & 277.67 & 2.37 & 266.70 & 17.64 & 489.03 & 30.01 & 382.82 & 5.83 & 476.44\\
    SAM-Lightening \cite{songa2024sam} & & 89.41 & 168.96 & 87.62 & 300.50 & 4.26 & 251.03 & 56.94 & 419.96 & 43.68 & 446.36 & 12.01 & 439.31\\
    \midrule 
    De-LightSAM & \ding{56} & \textbf{91.45} & \textbf{78.13} & \textbf{92.24} & \textbf{91.30} & \textbf{79.68} & \textbf{83.84} & \textbf{65.96} & \textbf{160.13} & \textbf{61.62} & \textbf{199.74} & \textbf{64.21} & \textbf{248.47}\\
    \bottomrule
    \end{tabular}}}
    \label{tab:2}
    \vspace{-1.0em}
\end{table*}

\begin{figure*}[!t]
  \centering
  \includegraphics[width=0.95\linewidth]{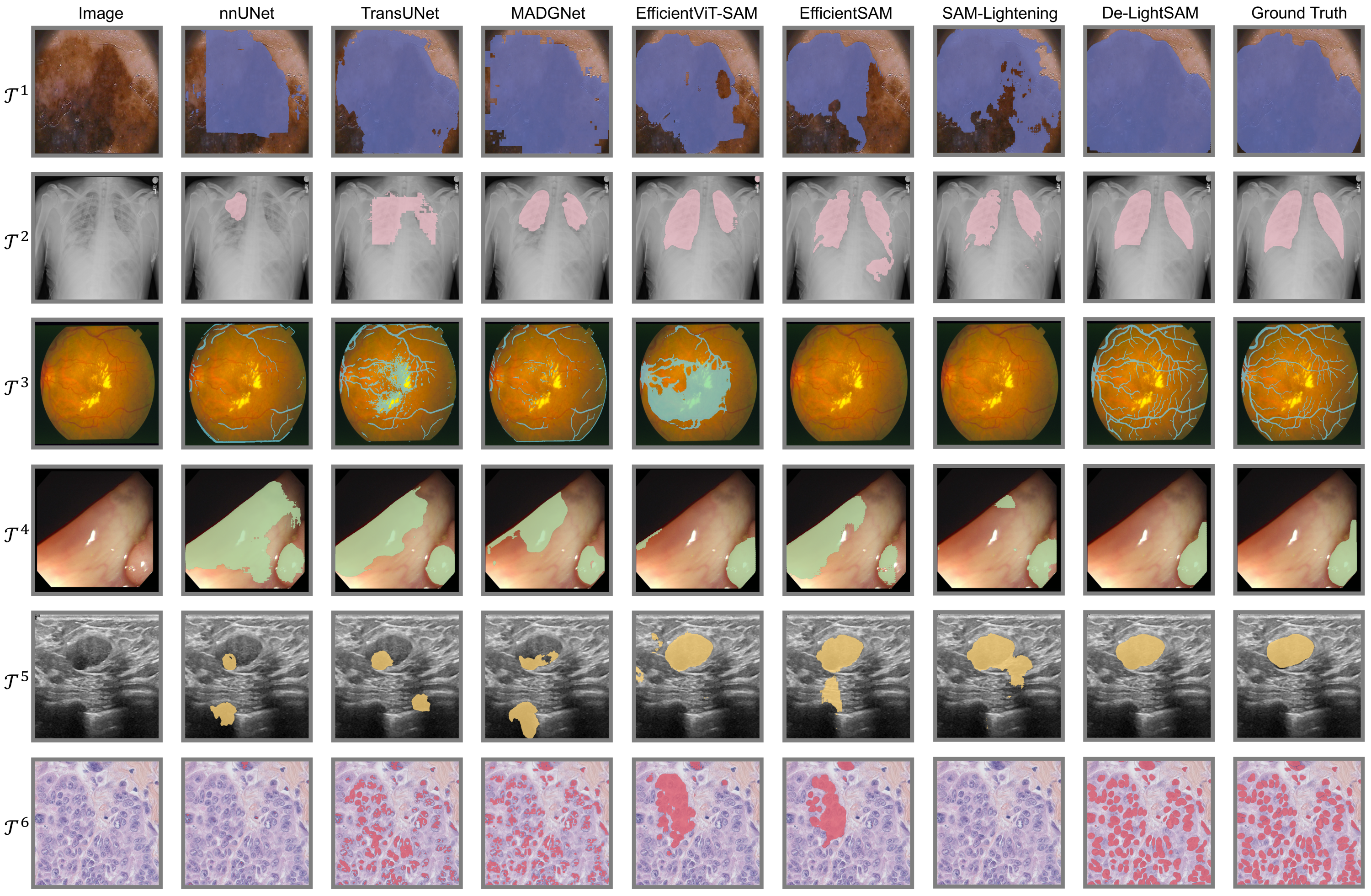}
  \vspace{-1.0em}
  \caption{Qualitative comparison with SOTA task-specific models and lightweight SAM frameworks on medical image segmentation (\textit{target domain}).}
  \label{fig:visual}
  \vspace{-1.0em}
\end{figure*}

\begin{table}[!t]
    \centering
    \small
    \caption{Comparison of Computation Costs with Automatic Segmentation Models under 1024 $ \times$ 1024 Inputs.}
    {\scalebox{0.91}{
    \begin{tabular}{lcccc}
    \toprule
    Method & Params $\downarrow$ & FLOPs $\downarrow$ & FPS $\uparrow$ & Latency $\downarrow$ \\
    \midrule
    U-Net \cite{ronneberger2015u} & 13.40M & 497.91G & 11.07 & 90.34ms\\
    U-Net++ \cite{zhou2019unet++} & 9.16M & 558.46G & 8.57 & 116.69ms\\
    Att-UNet \cite{schlemper2019attention} & 8.72M & 269.07G & 8.45 & 118.34ms\\
    nnUNet \cite{isensee2021nnu} & 34.29M & 554.40G & 10.99 & 91.00ms\\
    H2Former \cite{he2023h2former} & 33.71M & 536.96G & 7.41 & 134.95ms\\
    TransUNet \cite{chen2024transunet} & 109.54M & 865.31G & 4.73 & 211.42ms\\
    ADS \cite{xu2022adversarial} & 14.07M & 503.59G & 10.91 & 91.67ms\\
    CIAug \cite{ouyang2022causality} & 14.26M & 500.23G & 10.85 & 92.17ms\\
    MADGNet \cite{nam2024modality} & 31.40M & 210.65G & 8.33 & 120.05ms\\
    \midrule
    MobileSAM \cite{zhang2023faster}   & 9.79M & 39.71G & 1.05 & 952.38ms\\
    RepViT-SAM \cite{wang2023repvit} &  23.16M & 115.61G  & 1.12 & 892.86ms\\
    EfficientViT-SAM \cite{zhang2024efficientvit} & 34.80M & 89.10G & 0.91 & 1098.90ms\\
    EfficientSAM \cite{xiong2023efficientsam} & 25.38M & 32.51G & 0.98 & 1020.41ms\\
    EdgeSAM \cite{zhou2023edgesam}  & 9.60M & 22.10G & 1.25 & 800.00ms\\
    SAM-Lightening \cite{songa2024sam} & 19.26M & 52.46G & 1.02 & 980.39ms\\
    \midrule
     De-LightSAM  & 12.74M & 55.86G & 13.09 & 76.39ms\\
    \bottomrule
    \end{tabular}}}
    \label{tab:3}
    \vspace{-1.0em}
\end{table}

\begin{table}[!t]
    \centering
    \small
    \caption{Ablation study of De-LightSAM in domain-generalized Medical Image Segmentation: $\mathcal{S} \rightarrow \mathcal{T}$.}
    {\scalebox{0.91}{
    \begin{tabular}{lccc|cccc}
    \toprule
    Row & $M_1$ & $M_2$ & $M_3$ & Dice (Avg.) $\uparrow$ & HD (Avg.) $\downarrow$ & Params $\downarrow$\\
    \midrule
    1 & &  &  & 64.49 & 451.43 & 641.09M\\ 
    2 & \checkmark &  &  & 71.94 & 286.32 & 13.36M\\ 
    3 & & \checkmark &  & 69.96 & 152.65 & 641.11M\\
    4 & & & \checkmark & 67.28 & 297.51 & 640.45M\\
    5 & \checkmark & \checkmark & & 74.29 & 142.97 & 13.38M\\
    6 & \checkmark & & \checkmark & 75.17 & 272.88 & 12.72M\\
    7 & & \checkmark & \checkmark & 70.83 & 143.26 & 640.47M\\
    8 & \checkmark & \checkmark & \checkmark & \textbf{75.86} & \textbf{140.60} & 12.74M\\
    \bottomrule
    \end{tabular}}}
    \label{tab:7}
    \vspace{-1.0em}
\end{table}

\begin{table}[!t]
    \centering
    \small
    \setlength\tabcolsep{11pt}
    \caption{Ablation study of MDKD in domain-generalized Medical Image Segmentation: $\mathcal{S} \rightarrow \mathcal{T}$.}
    {\scalebox{1}{
    \begin{tabular}{ll|cccc}
    \toprule
    Row & Method & Dice (Avg.) $\uparrow$ & HD (Avg.) $\downarrow$ \\
    \midrule
    1 & DC-Encoder & 66.17 & 192.49\\
    2 & +KD  & 72.04 & 165.23 \\ 
    3 & +MDKD  & 75.86 & 140.60 \\
    \bottomrule
    \end{tabular}}}
    \label{tab:12}
    \vspace{-1.0em}
\end{table}

\section{Experiments}
\subsection{Datasets and Implementations}
\subsubsection{Datasets}

To validate the effectiveness of our proposed De-LightSAM, we select six different medical imaging modalities. Table \ref{tab:d1} presents the datasets of source domains for proving the universal ability. We adopt a common train-val-test split of 8:1:1 in PCXA, CVC-ClinicDB, UDIAT and DSB-2018. For ISIC-2018 \footnote{https://challenge.isic-archive.com/data/\#2018} and DRIVE \footnote{https://drive.grand-challenge.org/} datasets, we use the official train-val-test sets provided by their competition organizers. To evaluate the generalization capability of our framework, we further collect an unseen domain for each modality, as illustrated in Table \ref{tab:d2}. 

\subsubsection{Implementation Details}
We perform all experiments on a single NVIDIA A6000 GPU with PyTorch. We adopt the optimizer using Adam with a learning rate of $1\times10^{-4}$. The batch size and epochs are set to 4 and 200, respectively. We apply the exponential decay strategy to adjust the learning rate, where the factor is set as 0.98. All images are resized to $1024 \times 1024$ during the training and test stages. We set the patch size $m$ to be $32$, $L_2=1$ and utilize 1\% images of the SA-1B dataset for the distillation. We use ViT-H \cite{dosovitskiy2020image} as the teacher image encoder for all SAM-based architectures. The loss coefficient $\lambda$ is set to 0.7 during training. In comparing traditional automatic segmentation methods, we consider both UNet-based \cite{ronneberger2015u, zhou2019unet++, schlemper2019attention, isensee2021nnu, he2023h2former, chen2024transunet} and DG-based \cite{ouyang2022causality, xu2022adversarial, nam2024modality} fully-automated architectures as baselines. We reimplement these models, follow their official configurations in our experiments, and train a single model for each domain. In the comparison of lightweight SAM frameworks, MobileSAM \cite{zhang2023faster}, RepViT-SAM \cite{wang2023repvit}, EfficientViT-SAM \cite{zhang2024efficientvit}, EfficientSAM-Ti \cite{xiong2023efficientsam}, EdgeSAM \cite{zhou2023edgesam} and SAM-Lightening \cite{songa2024sam} are served as baselines. These architectures are trained with the point prompt mode that uses the \textit{ConnectedComponentsWithStats} function in OpenCV to calculate the centroid of each instance as point prompts. In addition to evaluating the manual point mode, they perform the automatic mask generation mode \cite{kirillov2023_sam}. 

\subsubsection{Evaluation Metrics}
To perform the comprehensive evaluation of medical image segmentation, we adopt two common metrics: Dice coefficient and Hausdorff Distance (HD). Both measure the similarity between the prediction and ground truth, where HD is more sensitive to the boundary than Dice. We also report model parameters, Floating Point Operations (FLOPs), Frames Per Second (FPS) and latency to reveal the computation cost and inference speed. To match the predicted $K$ masks with the corresponding ground truth $y^k_i$, we calculate the Dice score between $\{\hat{y}^k_i\}_{k=1}^K$ and $y^k_i$. The one with the highest Dice score in this set is chosen as the matching prediction mask for the evaluation \cite{huang2024segment}.

\subsection{Comparison with State-of-the-arts on Source Domains}

To validate the effectiveness of our lightweight De-LightSAM framework in universal medical segmentation, we compare it with SOTA fully-automated architectures: U-Net \cite{ronneberger2015u}, U-Net++ \cite{zhou2019unet++}, Att-UNet \cite{schlemper2019attention}, nnUNet \cite{isensee2021nnu}, H2Former \cite{he2023h2former}, TransUNet \cite{chen2024transunet}, ADS \cite{xu2022adversarial}, CIAug \cite{ouyang2022causality}, MADGNet \cite{nam2024modality} and lightweight SAMs \cite{zhang2023faster, wang2023repvit, zhang2024efficientvit, xiong2023efficientsam, zhou2023edgesam, songa2024sam} on the test set of source domains, as illustrated in Table \ref{tab:1}. For fair comparisons, the knowledge of lightweight SAMs is distilled from the same teacher model $\mathcal{F}_{\rm ViT-Med}$. Firstly, it can be observed that De-LightSAM performs better than all task-specific models, especially achieving a Dice of 85.28\% in $S^5$ and being 10.06\% higher than nnUNet \cite{isensee2021nnu}. Note that all task-specific models require to be retrained on each domain. Therefore, our De-LightSAM demonstrates outstanding universal capabilities. Secondly, compared to recent lightweight SAMs, the proposed De-LightSAM outperforms the laborious point-prompt mode. Particularly, with the automatic mask generation mode, the performance of these lightweight SAMs in retinal vessel and nuclei segmentation tasks declines rapidly. The proposed De-LightSAM significantly surpasses EfficientViT-SAM \cite{zhang2024efficientvit} with a 56.47\% and 25.92\% Dice increase, respectively. On HD metric, our framework reduces the distance by up to 5.8$\times$ compared to current SOTA models, revealing more accurate boundary prediction. These results reveal the superior universal capabilities of our De-LightSAM on diverse medical segmentation tasks without the demand for manual prompts. 
 
\subsection{Comparison with State-of-the-arts on Unseen Domains}
Furthermore, we evaluate the generalization capabilities of our De-LightSAM architecture on unseen target domains, which is provided in Table \ref{tab:2}. Within task-specific architectures, the large-capacity TransUNet \cite{chen2024transunet} and multi-task learning MADGNet \cite{nam2024modality} show better performance than nnUNet \cite{isensee2021nnu}. On the contrary, our method achieves overwhelming performance on all unseen domains with a significant rise of 1.74\%, 8.13\%, 12.80\%, 21.64\%, 16.71\% and 4.92\% over MADGNet on the Dice metric. Compared to the lightweight SAMs, the proposed De-LightSAM tackles the challenge of segmentation mask generations in retinal vessels, ultrasound cancer and microscopic nuclei segmentation tasks, improving the Dice score by more than 50\%. Remarkably, our De-LightSAM framework illustrates a lower HD distance, which provides a more precise localization for segmentation targets. Furthermore, we present the computation costs of each framework in Table \ref{tab:3}. In high-resolution medical image segmentation tasks, we observe that traditional task-specific architectures demonstrate acceptable inference speed but higher model complexity. The recent lightweight SAMs suffer from slow inference speed and high latency in the automatic mask generation mode due to the inefficient sliding window algorithm. In contrast, our De-LightSAM displays remarkable complexity-speed trade-offs. Overall, these results demonstrate the superior generalization capabilities of our proposed De-LightSAM framework on unseen domains across different medical imaging modalities with high computation efficiency. 

\begin{table}[!t]
    \centering
    \small
    \caption{Comparison with auto-prompting methods on domain-generalized medical image segmentation: $\mathcal{S} \rightarrow \mathcal{T}$.}
    {\scalebox{0.9}{
    \begin{tabular}{lccc}
    \toprule
    Methods & Prompt Types & Dice (Avg.) $\uparrow$ & HD (Avg.) $\downarrow$ \\
    \midrule
    CellSAM \cite{israel2024foundation} & Box & 74.51 & 149.73 \\
    UN-SAM \cite{chen2024sam} & LRMask & 73.87 & 145.59 \\
    UV-SAM \cite{zhang2024uv} & Box + LRMask & 74.24 & 148.14 \\
    PromptNucSeg \cite{shui2025unleashing} & Point & 73.95 & 153.72 \\
    SP-SAM \cite{wu2023self} & Box + Point & 73.39 & 157.63 \\
    \midrule
     De-LightSAM & Patch & \textbf{75.86} & \textbf{140.60} \\
    \bottomrule
    \end{tabular}}}
    \label{tab:8}
\end{table}

\begin{figure}[!t]
  \centering
  \includegraphics[width=1\linewidth]{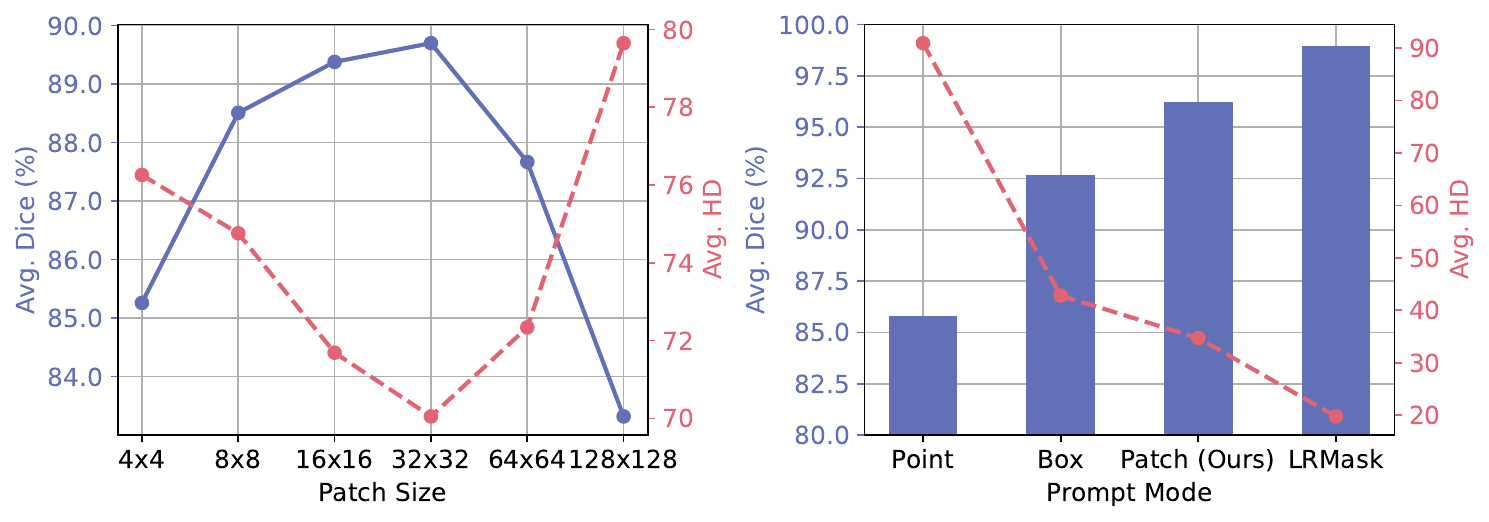}
   \vspace{-1.0em}
  \caption{Hyper-parameter analysis of patch size (left) and comparison of model performance based on different prompt types (right). Each prompt type utilizes manual annotations (ground truth) as the input.}
  \label{fig:diss}
  \vspace{-1.0em}
\end{figure}

\begin{figure}[!t]
  \centering
  \includegraphics[width=1\linewidth]{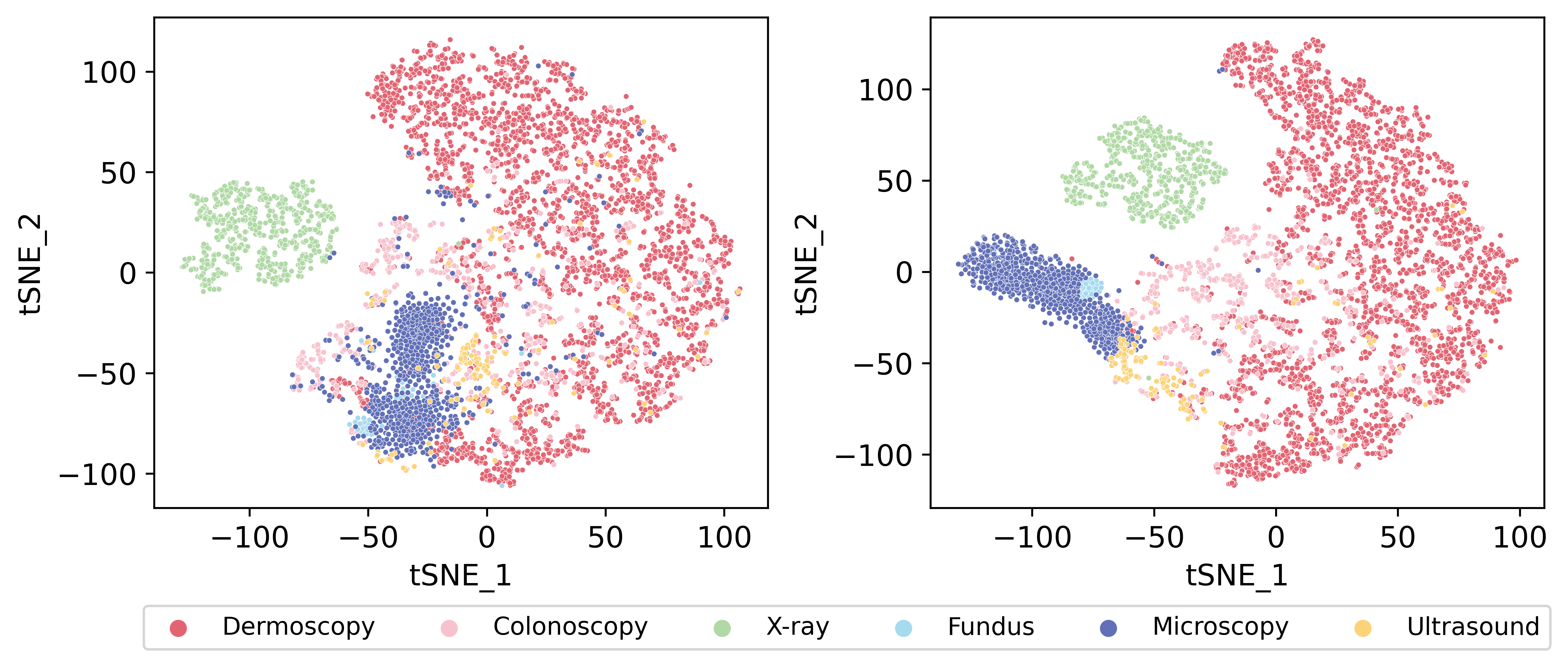}
   \vspace{-1.0em}
  \caption{Feature comparison via T-SNE between the baseline model (left) and our De-LightSAM framework (right) on unseen target domains.}
  \label{fig:tsne}
  \vspace{-1.0em}
\end{figure}

\subsection{Ablation Study}
To investigate the effectiveness of our proposed DC-Encoder $M_1$, SP-Generator $M_2$ and QM-Decoder $M_3$ modules, we further conduct a comprehensive ablation study on unseen target domains of six medical imaging modalities, displayed in Table \ref{tab:7}. By removing the devised modules from De-LightSAM, in $1^{st}$ row, the standard fine-tuned SAM \cite{kirillov2023_sam} serves as the ablation baseline. By separately adding the MDKD ($2^{nd}$ row), SP-Generator ($3^{rd}$ row) and QM-Decoder ($4^{th}$ row), the performance is increased with the average Dice of all modalities gain of 7.45\%, 5.47\%, 2.79\%, respectively. Particularly, the MDKD strategy decreases 67.90\% of the parameters compared to SAM. The introducing SP-Generator module ($3^{rd}$ row) significantly reduces HD by 66.19\%, which is more efficient than manual point prompts. The $5^{th}$ row to the $7^{th}$ row indicates the compatibility between each module. On this basis, our complete De-LightSAM framework ($8^{th}$ row) simultaneously adopts all three components to achieve the best performance across all modalities.

Furthermore, we conduct a detailed analysis of our MDKD strategy, as shown in Table \ref{tab:12}. We consider DC-Encoder trained from scratch as the baseline ($1^{st}$ Row), which achieves 66.17\% Dice and 192.49 HD. The addition of conventional knowledge distillation ($2^{nd}$ Row) improves performance to 72.04\% Dice with reduced HD of 165.23. However, our proposed MDKD strategy ($3^{rd}$ Row) demonstrates superior effectiveness, providing an additional 3.82\% Dice improvement and 24.63 HD reduction, validating the effectiveness of our decoupled distillation approach in handling diverse medical imaging scenarios. Overall, these ablation experiments prove that the DC-Encoder strategy significantly decreases the computation costs of De-LightSAM while maintaining competitive accuracy. SP-Generator eliminates the requirement for manual annotations. QM-Decoder further enhances the generalization capabilities of the model across diverse modalities by preventing cross-modal interference. The MDKD strategy enables effective knowledge transfer from foundation models to lightweight architectures, improving the generalization capabilities of De-LightSAM.  

\subsection{Discussion}

\subsubsection{Effectiveness of Patch Prompt Learning}
In this section, we delve into the rationale behind the adoption of the patch prompt learning strategy. Initially, we replace our SP-Generator module with other auto-prompting methods in De-LightSAM and make a comparison in Table \ref{tab:8}. It can be observed that Cell-SAM \cite{israel2024foundation}, UN-SAM\cite{zhang2024uv} and PromptNucSeg \cite{shui2025unleashing} perform better than UN-SAM \cite{chen2024sam} and SP-SAM \cite{wu2023self} but require more parameter consumptions as they utilize additional segmentors (\textit{e.g.,} Segformer \cite{xie2021segformer}) or detectors (\textit{e.g.,} DETR \cite{carion2020end}) to produce prompts. In contrast, our SP-Generator module eliminates these demands and outperforms state-of-the-arts by directly extracting high-quality patch prompts from the image embedding. In addition, we examine the efficiency of SP-Generator with different patch sizes on source domains. As illustrated in Fig. \ref{fig:diss} (left), based on the average performance on six modalities, setting a relatively large patch size proves beneficial for the patch category task, albeit it provides less semantic information than mask prompts. Conversely, reducing the patch size increases the prediction complexity, consequently leading to more errors or noise.
Additionally, we conduct a comparison of our devised patch prompt with point, box and Low-Resolution (LR) mask prompt modes in Fig. \ref{fig:diss} (right). Considering that automatic prompt generation aims to learn the representation of human annotations, all prompt types utilize manual annotations (ground truth) as inputs for the experiment. Observations reveal that the manual patch prompt outperforms both the box and point prompts. Therefore, the learning patch prompt approach proves more efficient for guiding the segmentation decoding.

\subsubsection{Significance of Modality-Decoupled Framework}
In the design of our MDKD strategy and image encoder, we adopt a decoupling strategy to provide an independent encoding workflow for each modality. To qualitatively evaluate the effectiveness of De-LightSAM in learning discriminative representations, we make a feature comparison via T-SNE with the baseline (using a full-knowledge sharing strategy). As shown in Fig. \ref{fig:tsne}, the features produced by our method exhibit significant discriminability over the six medical image modalities on unseen domains, which benefits the following decoder in terms of segmentation masks significantly, enhancing the universal and generalized capabilities of De-LightSAM.

\section{Conclusion}
In this paper, we have proposed the De-LightSAM framework for universal medical image segmentation. Specifically, the DC-Encoder has been introduced with a modality controller to generate discriminative features for diverse medical modalities while maintaining computational efficiency. Then, SP-Generator has been designed to automatically produce high-quality dense prompt embeddings for guiding segmentation decoding, eliminating the dependency on manual annotations. Moreover, QM-Decoder has established independent decoding channels for each modality, preventing cross-modal knowledge interference and enhancing modality-specific performance. Finally, the MDKD strategy has been devised to distill both common and domain-specific knowledge from foundation models to our lightweight architecture, unleashing generalization potentials. Extensive experiments have demonstrated that De-LightSAM achieves significantly lower computational complexity than the standard SAM and outperforms state-of-the-arts across diverse medical imaging segmentation tasks, exhibiting superior domain generalization capabilities.


\bibliographystyle{IEEEtran}
\bibliography{refs}

\begin{thebibliography}{10}
\providecommand{\url}[1]{#1}
\csname url@samestyle\endcsname
\providecommand{\newblock}{\relax}
\providecommand{\bibinfo}[2]{#2}
\providecommand{\BIBentrySTDinterwordspacing}{\spaceskip=0pt\relax}
\providecommand{\BIBentryALTinterwordstretchfactor}{4}
\providecommand{\BIBentryALTinterwordspacing}{\spaceskip=\fontdimen2\font plus
\BIBentryALTinterwordstretchfactor\fontdimen3\font minus \fontdimen4\font\relax}
\providecommand{\BIBforeignlanguage}[2]{{%
\expandafter\ifx\csname l@#1\endcsname\relax
\typeout{** WARNING: IEEEtran.bst: No hyphenation pattern has been}%
\typeout{** loaded for the language `#1'. Using the pattern for}%
\typeout{** the default language instead.}%
\else
\language=\csname l@#1\endcsname
\fi
#2}}
\providecommand{\BIBdecl}{\relax}
\BIBdecl

\bibitem{li2024focus}
H.~Li, D.-H. Zhai, Q.~Liu, K.~Tian, Y.~Yang, Z.~Chang, S.~Wang, and Y.~Xia, ``Focus-transunet3d: High-precision model for 3d segmentation of medical point targets,'' \emph{IEEE Trans. Circuits Syst. Video Technol.}, 2024.

\bibitem{zhou2024uncertainty}
T.~Zhou, Y.~Zhou, G.~Li, G.~Chen, and J.~Shen, ``Uncertainty-aware hierarchical aggregation network for medical image segmentation,'' \emph{IEEE Trans. Circuits Syst. Video Technol.}, 2024.

\bibitem{wang2024frequency}
D.~Wang, T.~Zhou, Y.~Zhang, S.~Gao, and J.~Yang, ``Frequency-aware interaction network for ultrasound image segmentation,'' \emph{IEEE Trans. Circuits Syst. Video Technol.}, 2024.

\bibitem{wu2023medsa_jinyueming}
J.~Wu, W.~Ji, Y.~Liu, H.~Fu, M.~Xu, Y.~Xu, and Y.~Jin, ``Medical sam adapter: Adapting segment anything model for medical image segmentation,'' \emph{arXiv preprint arXiv:2304.12620}, 2023.

\bibitem{ronneberger2015u}
O.~Ronneberger, P.~Fischer, and T.~Brox, ``U-net: Convolutional networks for biomedical image segmentation,'' in \emph{MICCAI}.\hskip 1em plus 0.5em minus 0.4em\relax Springer, 2015, pp. 234--241.

\bibitem{schlemper2019attention}
J.~Schlemper, O.~Oktay, M.~Schaap, M.~Heinrich, B.~Kainz, B.~Glocker, and D.~Rueckert, ``Attention gated networks: Learning to leverage salient regions in medical images,'' \emph{Med. Image Anal.}, vol.~53, pp. 197--207, 2019.

\bibitem{isensee2021nnu}
F.~Isensee, P.~F. Jaeger, S.~A. Kohl, J.~Petersen, and K.~H. Maier-Hein, ``nnu-net: a self-configuring method for deep learning-based biomedical image segmentation,'' \emph{Nature Methods}, vol.~18, no.~2, pp. 203--211, 2021.

\bibitem{he2023h2former}
A.~He, K.~Wang, T.~Li, C.~Du, S.~Xia, and H.~Fu, ``H2former: An efficient hierarchical hybrid transformer for medical image segmentation,'' \emph{IEEE Trans. Med. Imaging}, vol.~42, no.~9, pp. 2763--2775, 2023.

\bibitem{yun2023spectr}
B.~Yun, B.~Lei, J.~Chen, H.~Wang, S.~Qiu, W.~Shen, Q.~Li, and Y.~Wang, ``Spectr: Spectral transformer for microscopic hyperspectral pathology image segmentation,'' \emph{IEEE Trans. Circuits Syst. Video Technol.}, vol.~34, no.~6, pp. 4610--4624, 2023.

\bibitem{qiao2024effective}
Q.~Qiao, M.~Qu, W.~Wang, B.~Jiang, and Q.~Guo, ``Effective global context integration for lightweight 3d medical image segmentation,'' \emph{IEEE Trans. Circuits Syst. Video Technol.}, 2024.

\bibitem{li2023erdunet}
H.~Li, D.-H. Zhai, and Y.~Xia, ``Erdunet: An efficient residual double-coding unet for medical image segmentation,'' \emph{IEEE Trans. Circuits Syst. Video Technol.}, vol.~34, no.~4, pp. 2083--2096, 2023.

\bibitem{nam2024modality}
J.-H. Nam, N.~S. Syazwany, S.~J. Kim, and S.-C. Lee, ``Modality-agnostic domain generalizable medical image segmentation by multi-frequency in multi-scale attention,'' in \emph{CVPR}, 2024, pp. 11\,480--11\,491.

\bibitem{ouyang2022causality}
C.~Ouyang, C.~Chen, S.~Li, Z.~Li, C.~Qin, W.~Bai, and D.~Rueckert, ``Causality-inspired single-source domain generalization for medical image segmentation,'' \emph{IEEE Trans. Med. Imaging}, vol.~42, no.~4, pp. 1095--1106, 2022.

\bibitem{xu2022adversarial}
Y.~Xu, S.~Xie, M.~Reynolds, M.~Ragoza, M.~Gong, and K.~Batmanghelich, ``Adversarial consistency for single domain generalization in medical image segmentation,'' in \emph{MICCAI}.\hskip 1em plus 0.5em minus 0.4em\relax Springer, 2022, pp. 671--681.

\bibitem{dosovitskiy2020image}
A.~Dosovitskiy, L.~Beyer, A.~Kolesnikov, D.~Weissenborn, X.~Zhai, T.~Unterthiner, M.~Dehghani, M.~Minderer, G.~Heigold, S.~Gelly \emph{et~al.}, ``An image is worth 16x16 words: Transformers for image recognition at scale,'' in \emph{ICLR}, 2020.

\bibitem{kirillov2023_sam}
A.~Kirillov, E.~Mintun, N.~Ravi, H.~Mao, C.~Rolland, L.~Gustafson, T.~Xiao, S.~Whitehead, A.~C. Berg, W.-Y. Lo, P.~Dollar, and R.~Girshick, ``Segment anything,'' in \emph{ICCV}, October 2023, pp. 4015--4026.

\bibitem{huang2024segment}
Y.~Huang, X.~Yang, L.~Liu, H.~Zhou, A.~Chang, X.~Zhou, R.~Chen, J.~Yu, J.~Chen, C.~Chen \emph{et~al.}, ``Segment anything model for medical images?'' \emph{Med. Image Anal.}, vol.~92, p. 103061, 2024.

\bibitem{ma2024sam_majun_nc}
J.~Ma, Y.~He, F.~Li, L.~Han, C.~You, and B.~Wang, ``Segment anything in medical images,'' \emph{Nature Communications}, vol.~15, no.~1, p. 654, 2024.

\bibitem{na2024segment}
S.~Na, Y.~Guo, F.~Jiang, H.~Ma, and J.~Huang, ``Segment any cell: A sam-based auto-prompting fine-tuning framework for nuclei segmentation,'' \emph{arXiv preprint arXiv:2401.13220}, 2024.

\bibitem{zhang2024uv}
X.~Zhang, Y.~Liu, Y.~Lin, Q.~Liao, and Y.~Li, ``Uv-sam: Adapting segment anything model for urban village identification,'' in \emph{AAAI}, vol.~38, no.~20, 2024, pp. 22\,520--22\,528.

\bibitem{shui2025unleashing}
Z.~Shui, Y.~Zhang, K.~Yao, C.~Zhu, S.~Zheng, J.~Li, H.~Li, Y.~Sun, R.~Guo, and L.~Yang, ``Unleashing the power of prompt-driven nucleus instance segmentation,'' in \emph{ECCV}.\hskip 1em plus 0.5em minus 0.4em\relax Springer, 2025, pp. 288--304.

\bibitem{zhang2023faster}
C.~Zhang, D.~Han, Y.~Qiao, J.~U. Kim, S.-H. Bae, S.~Lee, and C.~S. Hong, ``Faster segment anything: Towards lightweight sam for mobile applications,'' \emph{arXiv preprint arXiv:2306.14289}, 2023.

\bibitem{xiong2023efficientsam}
Y.~Xiong, B.~Varadarajan, L.~Wu, X.~Xiang, F.~Xiao, C.~Zhu, X.~Dai, D.~Wang, F.~Sun, F.~Iandola \emph{et~al.}, ``Efficientsam: Leveraged masked image pretraining for efficient segment anything,'' \emph{CVPR}, 2024.

\bibitem{zhang2024efficientvit}
Z.~Zhang, H.~Cai, and S.~Han, ``Efficientvit-sam: Accelerated segment anything model without performance loss,'' \emph{arXiv preprint arXiv:2402.05008}, 2024.

\bibitem{zhou2019unet++}
Z.~Zhou, M.~M.~R. Siddiquee, N.~Tajbakhsh, and J.~Liang, ``Unet++: Redesigning skip connections to exploit multiscale features in image segmentation,'' \emph{IEEE Trans. Med. Imaging}, vol.~39, no.~6, pp. 1856--1867, 2019.

\bibitem{cai2024ultra}
Z.~Cai, Y.~Fan, M.~Zhu, and T.~Fang, ``Ultra-lightweight network for medical image segmentation inspired by bio-visual interaction,'' \emph{IEEE Trans. Circuits Syst. Video Technol.}, 2024.

\bibitem{zhao2024ultrasound}
X.~Zhao, Z.~Li, X.~Luo, P.~Li, P.~Huang, J.~Zhu, Y.~Liu, J.~Zhu, M.~Yang, S.~Chang \emph{et~al.}, ``Ultrasound nodule segmentation using asymmetric learning with simple clinical annotation,'' \emph{IEEE Trans. Circuits Syst. Video Technol.}, 2024.

\bibitem{zhang2023customized_sam_liudong}
K.~Zhang and D.~Liu, ``Customized segment anything model for medical image segmentation,'' \emph{arXiv preprint arXiv:2304.13785}, 2023.

\bibitem{mazurowski2023segment}
M.~A. Mazurowski, H.~Dong, H.~Gu, J.~Yang, N.~Konz, and Y.~Zhang, ``Segment anything model for medical image analysis: an experimental study,'' \emph{Med. Image Anal.}, vol.~89, p. 102918, 2023.

\bibitem{chen2024sam}
Z.~Chen, Q.~Xu, X.~Liu, and Y.~Yuan, ``Un-sam: Universal prompt-free segmentation for generalized nuclei images,'' \emph{arXiv preprint arXiv:2402.16663}, 2024.

\bibitem{hinton2015distilling}
G.~Hinton, O.~Vinyals, and J.~Dean, ``Distilling the knowledge in a neural network,'' \emph{arXiv preprint arXiv:1503.02531}, 2015.

\bibitem{yang2020knowledge}
J.~Yang, B.~Martinez, A.~Bulat, and G.~Tzimiropoulos, ``Knowledge distillation via softmax regression representation learning,'' in \emph{ICLR}, 2020.

\bibitem{zhao2022decoupled}
B.~Zhao, Q.~Cui, R.~Song, Y.~Qiu, and J.~Liang, ``Decoupled knowledge distillation,'' in \emph{CVPR}, 2022, pp. 11\,953--11\,962.

\bibitem{zhao2023efficient}
Q.~Zhao, L.~Zhong, J.~Xiao, J.~Zhang, Y.~Chen, W.~Liao, S.~Zhang, and G.~Wang, ``Efficient multi-organ segmentation from 3d abdominal ct images with lightweight network and knowledge distillation,'' \emph{IEEE Trans. Med. Imaging}, 2023.

\bibitem{li2022knowledge}
C.~Li, M.~Lin, Z.~Ding, N.~Lin, Y.~Zhuang, Y.~Huang, X.~Ding, and L.~Cao, ``Knowledge condensation distillation,'' in \emph{European Conference on Computer Vision}.\hskip 1em plus 0.5em minus 0.4em\relax Springer, 2022, pp. 19--35.

\bibitem{wu2022tinyvit}
K.~Wu, J.~Zhang, H.~Peng, M.~Liu, B.~Xiao, J.~Fu, and L.~Yuan, ``Tinyvit: Fast pretraining distillation for small vision transformers,'' in \emph{ECCV}.\hskip 1em plus 0.5em minus 0.4em\relax Springer, 2022, pp. 68--85.

\bibitem{zhou2023edgesam}
C.~Zhou, X.~Li, C.~C. Loy, and B.~Dai, ``Edgesam: Prompt-in-the-loop distillation for on-device deployment of sam,'' \emph{arXiv preprint arXiv:2312.06660}, 2023.

\bibitem{wang2023repvit}
A.~Wang, H.~Chen, Z.~Lin, J.~Han, and G.~Ding, ``Repvit-sam: Towards real-time segmenting anything,'' \emph{CVPR}, 2024.

\bibitem{songa2024sam}
Y.~Songa, B.~Pua, P.~Wanga, H.~Jiang, D.~Donga, and Y.~Shen, ``Sam-lightening: A lightweight segment anything model with dilated flash attention to achieve 30 times acceleration,'' \emph{arXiv preprint arXiv:2403.09195}, 2024.

\bibitem{howard2019searching}
A.~Howard, M.~Sandler, G.~Chu, L.-C. Chen, B.~Chen, M.~Tan, W.~Wang, Y.~Zhu, R.~Pang, V.~Vasudevan \emph{et~al.}, ``Searching for mobilenetv3,'' in \emph{ICCV}, 2019, pp. 1314--1324.

\bibitem{houlsby2019parameter}
N.~Houlsby, A.~Giurgiu, S.~Jastrzebski, B.~Morrone, Q.~De~Laroussilhe, A.~Gesmundo, M.~Attariyan, and S.~Gelly, ``Parameter-efficient transfer learning for nlp,'' in \emph{ICML}.\hskip 1em plus 0.5em minus 0.4em\relax PMLR, 2019, pp. 2790--2799.

\bibitem{tschandl2018ham10000}
P.~Tschandl, C.~Rosendahl, and H.~Kittler, ``The ham10000 dataset, a large collection of multi-source dermatoscopic images of common pigmented skin lesions,'' \emph{Sci. Data}, vol.~5, no.~1, pp. 1--9, 2018.

\bibitem{codella2019skin}
N.~Codella, V.~Rotemberg, P.~Tschandl, M.~E. Celebi, S.~Dusza, D.~Gutman, B.~Helba, A.~Kalloo, K.~Liopyris, M.~Marchetti \emph{et~al.}, ``Skin lesion analysis toward melanoma detection 2018: A challenge hosted by the international skin imaging collaboration (isic),'' \emph{arXiv preprint arXiv:1902.03368}, 2019.

\bibitem{candemir2013lung}
S.~Candemir, S.~Jaeger, K.~Palaniappan, J.~P. Musco, R.~K. Singh, Z.~Xue, A.~Karargyris, S.~Antani, G.~Thoma, and C.~J. McDonald, ``Lung segmentation in chest radiographs using anatomical atlases with nonrigid registration,'' \emph{IEEE Trans. Med. Imaging}, vol.~33, no.~2, pp. 577--590, 2013.

\bibitem{staal2004ridge}
J.~Staal, M.~D. Abr{\`a}moff, M.~Niemeijer, M.~A. Viergever, and B.~Van~Ginneken, ``Ridge-based vessel segmentation in color images of the retina,'' \emph{IEEE Trans. Med. Imaging}, vol.~23, no.~4, pp. 501--509, 2004.

\bibitem{bernal2015wm}
J.~Bernal, F.~J. S{\'a}nchez, G.~Fern{\'a}ndez-Esparrach, D.~Gil, C.~Rodr{\'\i}guez, and F.~Vilari{\~n}o, ``Wm-dova maps for accurate polyp highlighting in colonoscopy: Validation vs. saliency maps from physicians,'' \emph{Comput. Med. Imaging Graph.}, vol.~43, pp. 99--111, 2015.

\bibitem{yap2017automated}
M.~H. Yap, G.~Pons, J.~Marti, S.~Ganau, M.~Sentis, R.~Zwiggelaar, A.~K. Davison, and R.~Marti, ``Automated breast ultrasound lesions detection using convolutional neural networks,'' \emph{IEEE J. Biomed. Health Inform.}, vol.~22, no.~4, pp. 1218--1226, 2017.

\bibitem{caicedo2019nucleus}
J.~C. Caicedo, A.~Goodman, K.~W. Karhohs, B.~A. Cimini, J.~Ackerman, M.~Haghighi, C.~Heng, T.~Becker, M.~Doan, C.~McQuin \emph{et~al.}, ``Nucleus segmentation across imaging experiments: the 2018 data science bowl,'' \emph{Nature Methods}, vol.~16, no.~12, pp. 1247--1253, 2019.

\bibitem{mendoncca2013ph}
T.~Mendon{\c{c}}a, P.~M. Ferreira, J.~S. Marques, A.~R. Marcal, and J.~Rozeira, ``Ph 2-a dermoscopic image database for research and benchmarking,'' in \emph{EMBC}.\hskip 1em plus 0.5em minus 0.4em\relax IEEE, 2013, pp. 5437--5440.

\bibitem{tang2019xlsor}
Y.-B. Tang, Y.-X. Tang, J.~Xiao, and R.~M. Summers, ``Xlsor: A robust and accurate lung segmentor on chest x-rays using criss-cross attention and customized radiorealistic abnormalities generation,'' in \emph{MIDL}.\hskip 1em plus 0.5em minus 0.4em\relax PMLR, 2019, pp. 457--467.

\bibitem{hoover2003locating}
A.~Hoover and M.~Goldbaum, ``Locating the optic nerve in a retinal image using the fuzzy convergence of the blood vessels,'' \emph{IEEE Trans. Med. Imaging}, vol.~22, no.~8, pp. 951--958, 2003.

\bibitem{vazquez2017benchmark}
D.~V{\'a}zquez, J.~Bernal, F.~J. S{\'a}nchez, G.~Fern{\'a}ndez-Esparrach, A.~M. L{\'o}pez, A.~Romero, M.~Drozdzal, and A.~Courville, ``A benchmark for endoluminal scene segmentation of colonoscopy images,'' \emph{J. Healthc. Eng.}, vol. 2017, no.~1, p. 4037190, 2017.

\bibitem{al2020dataset}
W.~Al-Dhabyani, M.~Gomaa, H.~Khaled, and A.~Fahmy, ``Dataset of breast ultrasound images,'' \emph{Data in brief}, vol.~28, p. 104863, 2020.

\bibitem{naylor2018segmentation}
P.~Naylor, M.~La{\'e}, F.~Reyal, and T.~Walter, ``Segmentation of nuclei in histopathology images by deep regression of the distance map,'' \emph{IEEE Trans. Med. Imaging}, vol.~38, no.~2, pp. 448--459, 2018.

\bibitem{chen2024transunet}
J.~Chen, J.~Mei, X.~Li, Y.~Lu, Q.~Yu, Q.~Wei, X.~Luo, Y.~Xie, E.~Adeli, Y.~Wang \emph{et~al.}, ``Transunet: Rethinking the u-net architecture design for medical image segmentation through the lens of transformers,'' \emph{Med. Image Anal.}, p. 103280, 2024.

\bibitem{israel2024foundation}
U.~Israel, M.~Marks, R.~Dilip, Q.~Li, C.~Yu, E.~Laubscher, S.~Li, M.~Schwartz, E.~Pradhan, A.~Ates \emph{et~al.}, ``A foundation model for cell segmentation,'' \emph{bioRxiv}, pp. 2023--11, 2024.

\bibitem{wu2023self}
Q.~Wu, Y.~Zhang, and M.~Elbatel, ``Self-prompting large vision models for few-shot medical image segmentation,'' in \emph{MICCAI workshop on domain adaptation and representation transfer}.\hskip 1em plus 0.5em minus 0.4em\relax Springer, 2023, pp. 156--167.

\bibitem{xie2021segformer}
E.~Xie, W.~Wang, Z.~Yu, A.~Anandkumar, J.~M. Alvarez, and P.~Luo, ``Segformer: Simple and efficient design for semantic segmentation with transformers,'' \emph{NeurIPS}, vol.~34, pp. 12\,077--12\,090, 2021.

\bibitem{carion2020end}
N.~Carion, F.~Massa, G.~Synnaeve, N.~Usunier, A.~Kirillov, and S.~Zagoruyko, ``End-to-end object detection with transformers,'' in \emph{ECCV}.\hskip 1em plus 0.5em minus 0.4em\relax Springer, 2020, pp. 213--229.

\end{thebibliography}

\begin{IEEEbiography}[{\includegraphics[width=1in,height=1.25in,clip,keepaspectratio]{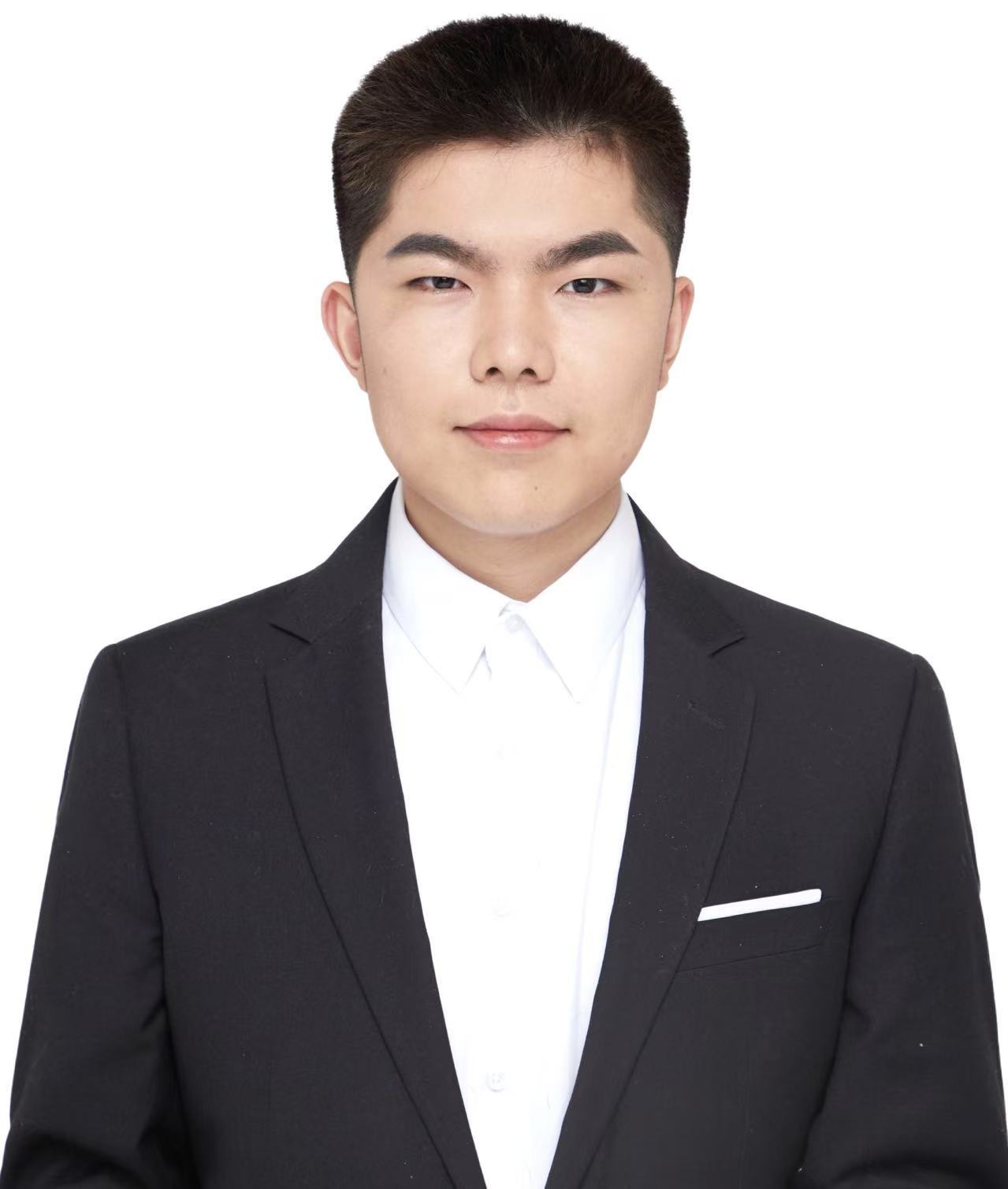}}]{Qing Xu}
received the B.Sc. degree from the University of Lincoln, Lincoln, UK, in 2021, the M.Sc. degree from the University of Hong Kong, Hong Kong SAR, in 2023. He is currently pursuing the Ph.D. degree with the University of Nottingham Ningbo China, Zhejiang, China. His research interests include medical image analysis, computer vision, and deep learning.
\end{IEEEbiography}

\begin{IEEEbiography}[{\includegraphics[width=1in,height=1.25in, clip,keepaspectratio]{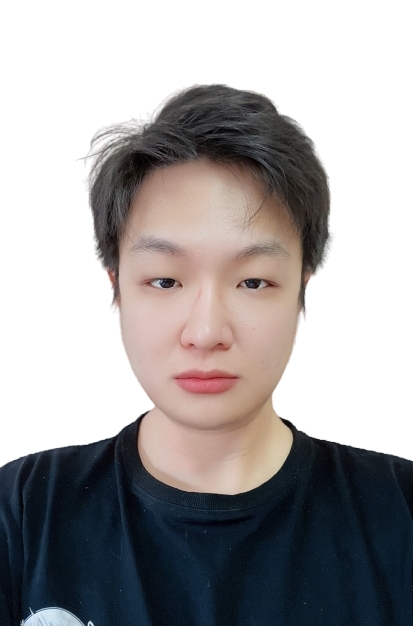}}]{Jiaxuan Li}received the B.Sc. degree from the University of New South Wales, Sydney, Australia, in 2021, the M.Sc. degree from the University of New South Wales, Sydney, Australia, in 2022. He is currently pursuing the Ph.D. degree with the University of Nottingham Ningbo China, Zhejiang, China. His research interests include medical image analysis, self-supervised Learning, computer vision, and deep learning.
\end{IEEEbiography}

\begin{IEEEbiography}[{\includegraphics[width=1in,height=1.25in, clip,keepaspectratio]{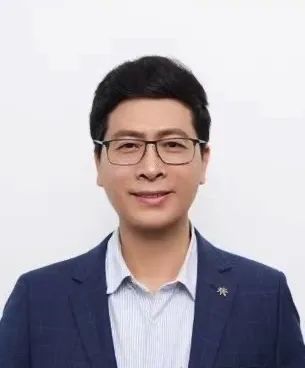}}]{Xiangjian He} (Senior Member, IEEE) received his Ph.D. degree from the University of Technology Sydney, Australia, in 1999. He is currently the Chair Professor of Computer Science and Technology and the Director of the Computer Vision and Intelligent Perception Laboratory in the University of Nottingham Ningbo China. His research interests include image processing, pattern recognition, computer vision, medical image segmentation and classification, and machine learning.
\end{IEEEbiography}

\begin{IEEEbiography}[{\includegraphics[width=1in,height=1.25in, clip,keepaspectratio]{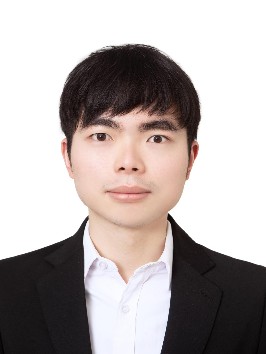}}]{Chenxin Li} received the B.Sc. degree from Xiamen University, China in 2019, the M.Sc. degree from the Xiamen University, China in 2022. He is currently pursuing the Ph.D. degree with the The Chinese University of Hong Kong, Hongkong SAR. His research interests include medical image analysis, computer vision, and deep learning.
\end{IEEEbiography}

\begin{IEEEbiography}[{\includegraphics[width=1in,height=1.25in, clip,keepaspectratio]{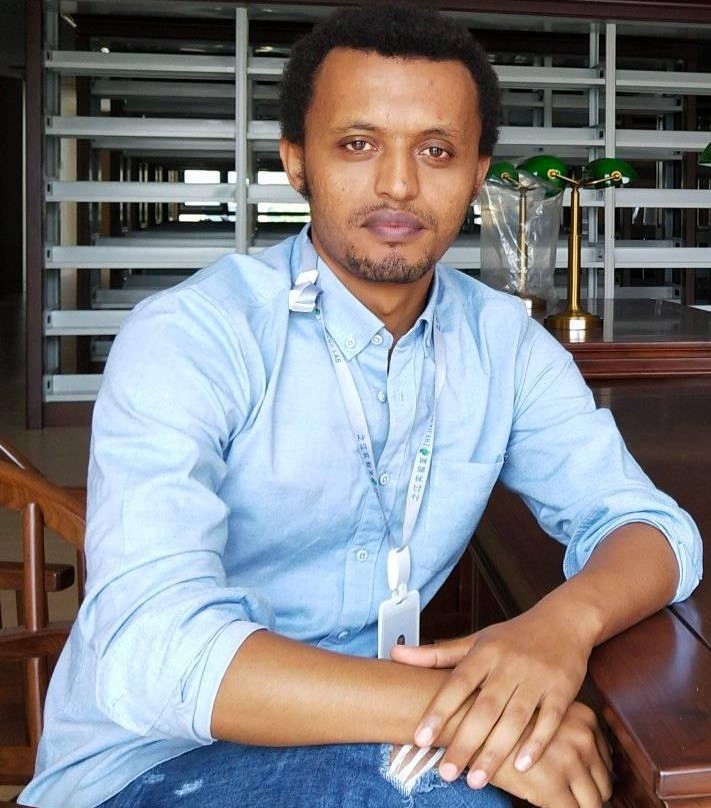}}]{Fiseha Berhanu Tesema} received his Ph.D. from the University of Electronic Science and Technology of China in 2020. Since May 2023, he has been serving as an Assistant Professor at the School of Computer Science, Faculty of Science and Engineering, University of Nottingham Ningbo China. His research interests include medical image analysis, artificial intelligence, visual perception, computer vision, machine and deep learning, human–robot interaction, and multimodal fusion.
\end{IEEEbiography}

\begin{IEEEbiography}[{\includegraphics[width=1in,height=1.25in, clip,keepaspectratio]{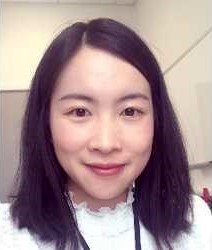}}]{Wenting Duan} received the B.Eng. and Ph.D. degrees in Electronic Engineering from the University of Sheffield, U.K., in 2006 and 2011, respectively. Since January 2012, she has been with the School of Engineering and Physical Sciences, University of Lincoln, U.K., where she is currently a Senior Lecturer in Computer Science. Her research interests include medical image analysis, multimodal fusion, image segmentation, and object detection.
\end{IEEEbiography}

\begin{IEEEbiography}[{\includegraphics[width=1in,height=1.25in, clip,keepaspectratio]{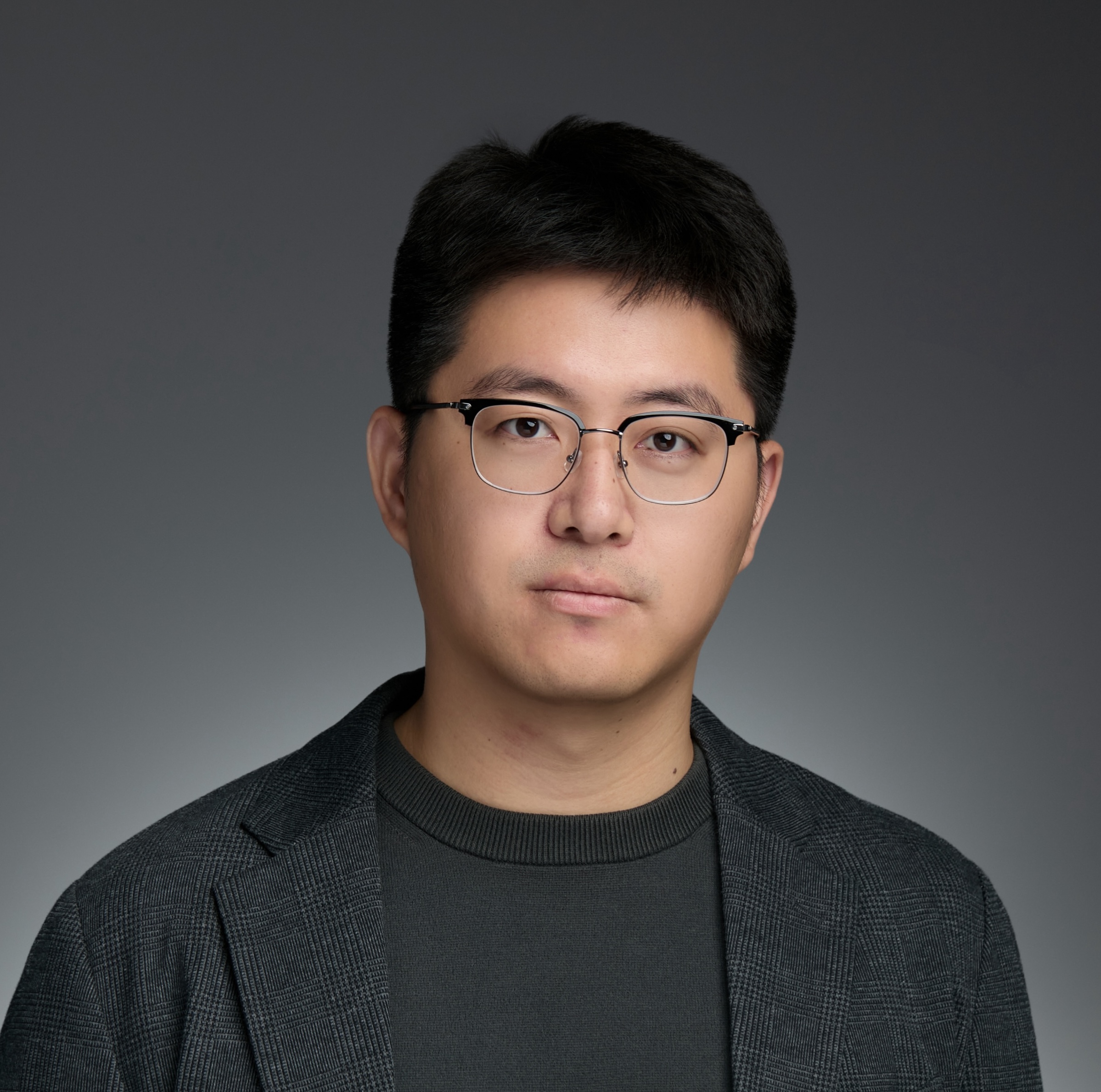}}]{Zhen Chen} received the B.Eng. from Xi'an Jiaotong University in 2016, the M.Eng. from University of Science and Technology of China in 2019, and the Ph.D. degree in Electrical Engineering from City University of Hong Kong in 2022. Dr. Chen worked as an assistant professor in Hong Kong Institute of Science and Innovation, Chinese Academy of Sciences. His research includes computer-aided diagnosis, computer vision and machine learning.
\end{IEEEbiography}

\begin{IEEEbiography}[{\includegraphics[width=1in,height=1.25in, clip,keepaspectratio]{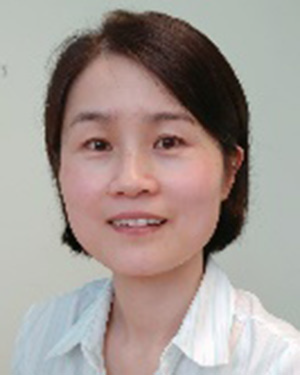}}]{Rong Qu} (Senior Member, IEEE) received the Ph.D. degree in computer science from the University of Nottingham, Nottingham, U.K., in 2003. She is a Professor with the University of Nottingham, Nottingham, U.K. Her main research interests include the modeling and evolutionary computation in combinatorial optimization using cross-disciplinary techniques, such as automated algorithms, hyper-heuristics, computational
intelligence, machine learning, constraint programming, and knowledge-based systems. 
\end{IEEEbiography}

\begin{IEEEbiography}[{\includegraphics[width=1in,height=1.25in, clip,keepaspectratio]{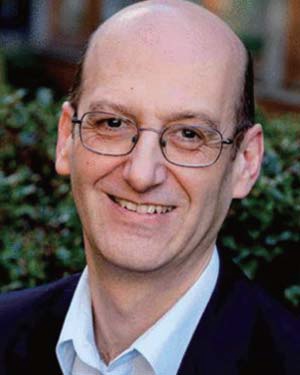}}]{Jonathan M. Garibaldi} (Fellow, IEEE) received the Ph.D. degree from the University of Plymouth, Plymouth, U.K, in 1997. He is currently the Provost of the University of Nottingham Ningbo, China. He has authored/coauthored more than 200 articles on fuzzy systems and intelligent data analysis. His research interests include modeling uncertainty and variation in human reasoning and in modeling medical domains.
\end{IEEEbiography}

\begin{IEEEbiography}[{\includegraphics[width=1in,height=1.25in, clip,keepaspectratio]{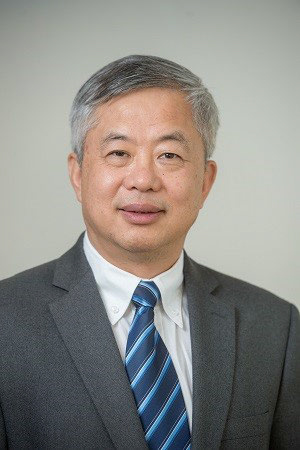}}]{Changwen Chen} (Fellow, IEEE) received the the Ph.D. degree from the University of Illinois at Urbana–Champaign in 1992. He is currently the Chair Professor of visual computing with
The Hong Kong Polytechnic University. His research
interests include multimedia communication, multimedia systems, mobile
video streaming, the Internet of Video Things (IoVT), image/video processing,
computer vision, deep learning, multimedia signal processing, and immersive
mobile video.

\end{IEEEbiography}

\vfill

\end{document}